\documentclass[showpacs,preprintnumbers,twocolumn,reprint,superscriptaddress,nobibnotes,nofootinbib]{revtex4-1}

\usepackage{amsmath,amssymb,hyperref,times}
\usepackage{graphicx}

\usepackage{dcolumn}
\def\<{\langle}
\def\>{\rangle}
\newcommand{\JSTAT}{J.~Stat.~Mech.}

\newcommand{\PRL}{Phys.~Rev.~Lett.}

\newcommand{\PRE}{Phys.~Rev. E}
\newcommand{\PRX}{Phys.~Rev. X}

\newcommand{\JPCM}{J.~Phys.: Condens.~Matter}
\newcommand{\JPAOLD}{J.~Phys.~A: Math.~Gen.}

\newcommand{\physrep}{Phys.~Rep.}
\newcommand{\EPLOLD}{Europhys. Lett.}
\newcommand{\EPJB}{Eur.~Phys.~J. B}
\newcommand{\ZPB}{Z.~Phys.~B}
\newcommand{\IJMPB}{Int.~J.~Mod.~Phys. B}

\newcolumntype{.}{D{.}{.}{-1}}

\begin{document}

\title{Critical behavior in the presence of an order-parameter pinning field}

\author{\firstname{Francesco} \surname{Parisen Toldin}}
\email{francesco.parisentoldin@physik.uni-wuerzburg.de}
\author{\firstname{Fakher F.} \surname{Assaad}}
\email{assaad@physik.uni-wuerzburg.de}
\affiliation{\mbox{Institut f\"ur Theoretische Physik und Astrophysik, Universit\"at W\"urzburg, Am Hubland, D-97074 W\"urzburg, Germany}}
\author{\firstname{Stefan} Wessel}
\email{wessel@physik.rwth-aachen.de}
\affiliation{Institut f\"ur Theoretische Festk\"orperphysik, JARA-FIT and JARA-HPC, RWTH Aachen University, D-52056 Aachen, Germany}
\begin{abstract}
We apply a recently advocated simulation scheme that employs a local order-parameter pinning field to study quantum critical phenomena in the two-dimensional square-lattice bilayer quantum Heisenberg model. Using a world-line quantum Monte Carlo approach, we show that for this model, the pinning-field approach allows to locate the quantum critical point over a wide range of pinning-field strengths. However, the identification of the quantum critical scaling behavior is found to be hard since the pinning field introduces strong corrections to scaling. In order to further elucidate the scaling behavior in this situation, we also study an improved classical lattice model in the three-dimensional Ising universality class by means of Monte Carlo simulations on large lattice sizes, which allow us to employ refined finite-size scaling considerations. A renormalization group analysis exhibits the presence of an important crossover effect from the zero pinning-field to a critical adsorption fixed point. In line with field-theoretical results, we find that at the critical adsorption fixed point the short-distance expansion of the order-parameter profile exhibits a new universal critical exponent. This result also implies the presence of slowly decaying scaling corrections, which we analyze in detail.
\end{abstract}

\maketitle

\section{Introduction}

The study of quantum phase transitions takes a central role in contemporary condensed matter physics. Of particular interest in many situations are quantum phase transitions towards phases with unconventional order parameters or into phases of strongly-interacting matter that exhibit no conventional ordering at all, such as in quantum spin liquid phases. For an unbiased numerical study of such systems, it is thus important to be able to resolve even weakly developed order parameters and their scaling behavior in the vicinity of the quantum phase transition. 
Recently, a new approach has been put forward for detecting spontaneous symmetry breaking in numerical simulations, where traditional approaches, based on order-parameter correlation functions, may be limited due to exceedingly small values of the order parameter in the thermodynamic limit~\cite{AH-13}. In particular, it was proposed to enhance the power of finite-size based studies by turning on a local ordering field, conjugate to the order parameter, applied, e.g., in a lattice model at a single site in a finite system, and to monitor the effect of this order-parameter pinning field over large distances from the coupling site. In addition to assessing the size of the order parameter in the thermodynamic limit, this approach was also suggested  for the study of quantum critical phenomena, in particular to determine quantum critical points and the associated critical exponents.
The introduction of symmetry-breaking terms in the Hamiltonian as a tool to investigate its phases is common also to Density Matrix Renormalization Group studies~\cite{WC-07} and to lattice QCD~\cite{BU-16}. 
A related means of investigating (quantum critical) spin systems is provided by the response to impurity spins or vacancies. This approach has been intensively examined  by both analytical and numerical approaches, see, e.g., Refs.~[\onlinecite{impi1,impi2,impi3,impi4}].

While the pinning-field approach has been used already for a study of fermionic systems~\cite{AH-13}, here we apply this method to the spin-1/2 bilayer Heisenberg model, in order to  assess the ability of the pinning-field method to (i) locate the quantum phase transition point and (ii) to estimate the values of underlying critical exponents. In particular, since the bilayer system's critical properties are well characterized and conventional simulation schemes, based on order-parameter correlation functions, were found to provide  accurate results at criticality, we consider this system an ideal test-case for the pinning-field approach, for which we can access finite lattices of comparable size and also extend beyond the size-limitations of, e.g., determinantal quantum Monte Carlo schemes.  

As will be discussed in  detail below, the presence of a finite pinning field constitutes a relevant perturbation (in the renormalization group sense) to the zero-field fixed point, even though it is applied  to (only) a single site on the lattice. This requires a refined finite-size scaling analysis beyond the conventional leading finite-size scaling ansatz for a reliable extraction of the critical exponents of the underlying quantum critical point. In order to quantify these scaling corrections on larger linear system sizes than those accessible by  quantum Monte Carlo approaches, we also consider here an improved classical lattice model in the presence of a pinning-field line, motivated by the general quantum-to-classical mapping.  This improved classical model (in the sense that scaling corrections in the absence of the pinning field are suppressed) in fact turns out to be an interesting model in its own, representing a magnetic defect line in a critical three-dimensional Ising system. Such a situation has indeed been analyzed previously, and we contribute here to the analysis of such defect-line systems, in particular by exhibiting the presence of a new universal critical exponent, which characterizes the order-parameter profile in the vicinity of the defect-line. 

Our Monte Carlo-based numerical studies are complemented by an extended analysis of  scaling relations in the presence of an order-parameter pinning field (a defect-line in the classical case) as well as a theoretical treatment based on renormalization group and short-distance-expansion considerations. These exhibit an important crossover  from the zero pinning-field fixed point to a critical adsorption fixed point that corresponds to an infinite pinning-field strength. In line with previous field-theoretical results, we find that the short-distance expansion of the order-parameter profile exhibits a new universal critical exponent at the critical adsorption fixed point.
This result also implies the presence of  slowly decaying scaling corrections in the order-parameter scaling function, which we analyze in detail and compare to our numerical findings in both the improved classical model and the bilayer quantum Heisenberg spin model.

In this context, we remark that critical adsorption is a well-known phenomenon in fluids which has attracted numerous experimental studies (see, e.g., Ref.~[\onlinecite{Law-01}] for a review, and Refs.~[\onlinecite{CL-01,JLPZ-01,CL-01b,CLG-02,CL-05,BZCMWS-05,BLSHWCM-07}] for more recent results), as well as theoretical investigations (see, e.g., Refs.[\onlinecite{BE-85,Cardy-90,CD-90,DC-91,DS-93,*DS-93_erratum,Diehl-94,FD-95,HD-99,HKSD-99,Hanke-00,PD-04}]).
In fact, as discussed in the conclusions, for the case of the classical model in the Ising universality class the present setup can be experimentally realized in the physics of fluids.

The remainder of the paper is organized as follows. In Sec.~\ref{sec:models}, we define in detail the model systems that we will analyze in our Monte Carlo studies. Then, in Sec.~\ref{sec:fss}, we provide a throughout analysis of the finite-size scaling behavior in the presence of a pinning field. This also includes an extended discussion of the order-parameter profile and the local susceptibility in the vicinity of the pinning-field center as well as the dominant corrections to scaling induced by the pinning field. Based on this theoretical considerations, we then analyze in Sec. ~\ref{sec:results} the numerical Monte Carlo results for both the classical lattice model as well as the bilayer quantum Heisenberg model. 
Finally, in Sec.~\ref{sec:conclusions}, we summarize our results and discuss future directions.
Appendix \ref{sec:chiloc} contains an alternative calculation of the renormalization-group dimension of the pinning field at the zero-field fixed point, which complements the determination presented in Sec.~\ref{sec:fss:h0}. In Appendix \ref{sec:honeycomb}, we discuss the implications of the present study for the critical behavior of the Hubbard model on the honeycomb lattice; this model was extensively studied with the pinning-field approach in  Ref.~[\onlinecite{AH-13}]. In Appendix \ref{sec:Sn}, we provide a rigorous argument concerning the magnetization scaling exponent for a classical lattice model in the presence of a pinning field coupled to a single site; such a setup is discussed in Sec.~\ref{sec:conclusions} as a possible generalization of the models studied here.

\section{Models}\label{sec:models}
In this section, we  introduce  the two lattice models that we use to analyze specifically the effects of pinning-field defects in (i) a two-dimensional quantum critical magnet and  (ii) the magnetic defect-line in a three-dimensional critical Ising system, respectively. 

\subsection{Bilayer quantum Heisenberg model}
\label{sec:models:heisenberg}
For the analysis of the effect of a local pinning field on a quantum critical magnetic system, we consider the spin-1/2 Heisenberg model on a square lattice bilayer. This model is described by the Hamiltonian
\begin{equation}
H = J \sum_{\alpha=1,2}\sum_{\<i\ j\>}\vec{S}_{i,\alpha}\vec{S}_{j,\alpha} + J' \sum_i\vec{S}_{i,1}\vec{S}_{i,2}.
\label{HHeisenberg}
\end{equation}
Here, we consider two layers of parallel stacked square lattices, with $\vec{S}_{i,\alpha}$ a spin-1/2 degree of freedom on the $i$th lattice site on the upper ($\alpha=1$) and lower ($\alpha=2$) layer. The first term  in $H$ accounts for a nearest-neighbor antiferromagnetic ($J>0$) spin exchange interaction within each layer, while the second term accounts for a local antiferromagnetic ($J'>0$) coupling between adjacent sites from each layer.  In the following, we denote by $g=J'/J$ the coupling ratio between the inter- and intralayer coupling strengths. 

This bilayer Heisenberg model exhibits a  quantum phase transition between a collinear antiferromagnetic phase at small values of $g$ and a large-$g$ quantum disordered dimer paramagnet, with a quantum critical point located at 
the critical ratio~\cite{Wang06} $g_c=2.5220(1)$.
The model has been analyzed  intensively in the past~\cite{Singh88,Millis93,Millis94,Sandvik94,Chubukov95,Wang06}, and in particular large-scale quantum Monte Carlo (QMC) simulations have both identified the above quoted location of the quantum critical point and verified the three-dimensional O(3) universality class (UC) of the quantum phase transition. Due to the absence of geometric frustration, this system can be studied using QMC without any sign problem, even in close vicinity of the quantum critical point, using by-now standard cluster update algorithms~\cite{Sandvik99,Syljuasen02,Alet05}.

Conventionally, in order to detect antiferromagnetic order on a bipartite lattice model with an SU(2) invariant Hamiltonian $H$, an estimator for the antiferromagnetic order parameter $m$ based on the spin-spin correlation functions is employed,
\begin{equation}
\label{eq:mconv}
 m(L)=\sqrt{\frac{1}{N^2}\sum_{i,j=1}^{N}\epsilon_i\epsilon_j \< {\bf S}_i\cdot {\bf S}_j\>},
\end{equation}
where $L$ denotes the linear system size, $N$ the number of lattice sites ($N=2L^2$ for the bilayer model under consideration here), and $\epsilon_i=\pm 1$, depending on the sublattice to which lattice site $i$ belongs. The finite-size data $m(L)$ then need to be extrapolated to the thermodynamic limit (TDL) in order to obtain the TDL value of the order parameter $m$. In cases, where $m$ becomes exceedingly small, which will be the case, e.g., when locating the system close to the critical point, the authors of Ref.~[\onlinecite{AH-13}] propose instead to modify the system's Hamiltonian by adding a symmetry-breaking  field term that couples to the order-parameter field. In contrast to the conventional symmetry-breaking field procedure, however, they propose to add a {\it local} term [given explicitly here for the case of a (antiferro-) magnetic transition]
\begin{equation}
H_{\mathrm{loc}}=h_0 S^z_{i_0}
\label{pf_interaction}
\end{equation}
in the Hamiltonian, where $h_0$ sets the magnitude of this local pinning field coupled to the spin on lattice site $i_0$, and which induces (antiferro-) magnetic correlations in the system by pinning the ordering direction. A QMC estimator for the order parameter $m$ is then based on evaluating the spatially averaged pinned order-parameter field,
\begin{equation}
\label{eq:mnew}
m(L)=\frac{1}{N} \sum_{i=1}^N \epsilon_i \< S^z_i\>_{h_0},
\end{equation}
again with a final extrapolation to the TDL.  Here, $\< ...\>_{h_0}$ indicates that the expectation value is to be taken for a system with Hamiltonian  $H+H_{\mathrm{loc}}$. This estimator will be dominated for large system sizes by the scaling of the pinned expectation value $\< S^z_i\>_{h_0}$ at large distances from the pinning site (which defines here the lattice site $i_0$). 
For a translational invariant Hamiltonian $H$, the pinning center site $i_0$ can of course be chosen arbitrary.  We refer to Ref.~[\onlinecite{AH-13}]  for a more detailed exposition of the pinning-field approach. For this paper, we  have applied this method to the bilayer Heisenberg model and examined its performance in locating the quantum phase transition and determine the critical properties, in particular the scaling behavior near criticality. 

\subsection{Classical lattice model}
\label{sec:models:bc}
In order to elucidate further the scaling behavior in the presence of a pinning field, we  also considered a three-dimensional classical lattice model, in the presence of a local magnetic field, acting on a single defect line. In fact, as discussed in Sec.~\ref{sec:fss}, under the quantum-to-classical mapping the model defined in Eq.~(\ref{HHeisenberg}) becomes equivalent to a three-dimensional classical model in the presence of a magnetic field restricted to a line, parallel to the imaginary time axis.
We study the Blume-Capel model~\cite{Blume-66,Capel-66}, which is a lattice spin model where the spin variables take values $S=-1,0,1$. It is defined by the classical reduced Hamiltonian
\begin{equation}
\begin{split}
{\cal H}&=-K\sum_{\left\< \substack{x,y,z\\ x',y',z'}\right\>}S_{(x,y,z)} S_{(x',y',z')} + \Delta\sum_{x,y,z} S_{(x,y,z)}^2 \\
&- h_0 \sum_{z}S_{(0,0,z)} - h\sum_{x,y,z} S_{(x,y,z)},\quad S_{(x,y,z)}=-1,0,1,
\end{split}
\label{HBC}
\end{equation}
such that the Gibbs weight is $\exp(-\cal H)$. The model is defined on a three-dimensional simple cubic lattice of size $L\times L\times L_z$, with periodic boundary conditions, where each site has Cartesian coordinates $(x,y,z)$.
In Eq.~(\ref{HBC}), the first sum extends over the nearest-neighbors pairs, the second and the last one over all the lattice sites. The third sum in Eq.~(\ref{HBC}) represents the pinning field and extends over a line parallel to the $z$ axis. In order to exploit the translational invariance of the model in the pinning-field direction, it is convenient to fix the origin of the coordinate system on a lattice site on the pinning field, such that a lattice site vector $\vec{x}$ decomposes into two components, parallel $\vec{x}_\parallel=z\hat{z}$ and perpendicular $\vec{x}_\perp=(x,y)$ to the pinning-field line, which is located at $\vec{x}_\perp=0$. Besides the pinning field $h_0$, we also consider the effect of a bulk field $h$.

In line with the convention used in previous works~\cite{Hasenbusch-10,Hasenbusch-10c,PTD-10,Hasenbusch-11,PTTD-13,PTTD-14,Hasenbusch-14,PT-13}, in the following, we fix the parameter $\Delta$, treating it as a part of the integration measure over the spin configuration, and vary the remaining parameters $K$, which controls the distance to the critical point, $h$, and $h_0$. The Blume-Capel model reduces to the usual Ising model in the limit $\Delta\rightarrow-\infty$. In the $(K,\Delta)$ plane, it exhibits a second-order transition line in the Ising UC, which extends from $\Delta=-\infty$ to the tricritical point $\Delta_{\rm tri}$. In dimension three $\Delta_{\rm tri}$ has been determined as $\Delta_{\rm tri}=2.006(8)$~\cite{Deserno-97,ZFJ-15}, as $\Delta_{\rm tri}\simeq 2.05$~\cite{HB-98}, and as $\Delta_{\rm tri}=2.0313(4)$~\cite{DB-04}. At $\Delta=0.656(20)$~\cite{Hasenbusch-10} the leading scaling corrections $\propto L^{-\omega}$, with $\omega=0.832(6)$~\cite{Hasenbusch-10} are suppressed and the model is ``improved''~\cite{PV-02}. As in recent numerical studies which employ this model~\cite{Hasenbusch-10c,PTD-10,Hasenbusch-11,PTTD-13,PTTD-14,Hasenbusch-14,PT-13}, here we fixed $\Delta=0.655$. At this value of $\Delta$, the model is critical for $K=K_c=0.387721735(25)$~\cite{Hasenbusch-10}. An accurate estimate of the critical exponents of the three-dimensional Ising UC, $\nu=0.63002(10)$ and $\eta=0.03627(10)$, has been obtained by using the improved Blume-Capel model~\cite{Hasenbusch-10}.

\section{Finite-Size Scaling}
\label{sec:fss}
\subsection{General properties}
\label{sec:fss:general}
Under the quantum-to-classical mapping, the model of Eq.~(\ref{HHeisenberg}) becomes equivalent to a model in cylindrical geometry with $D=d+1=3$ dimensions, where the inverse temperature $1/T$ corresponds to the size in imaginary time.
This is accomplished by expressing the partition function as a path integral in a $(d+1)$-dimensional space, using the base of spin coherent states~\cite{Sachdev-book}.
The presence of a pinning field gives rise to a line defect, parallel to the imaginary time axis, where the pinning field $h_0$ is coupled to. 
In contrast to the case of a vacancy~\cite{impi1,impi2},  there is no Berry phase term entering the field-theoretical description of the perturbation introduced to the bulk system by the pinning field.
While in general the scaling behavior of the inverse temperature is characterized by a nontrivial exponent $z$ and an anisotropic scaling~\cite{Sachdev-book}, $O(N)$-symmetric models exhibits a full Lorentzian symmetry at the critical point, with $z=1$~\cite{CSY-94}.

Here and in the following, we shall discuss the finite-size scaling (FSS) behavior, which we use to analyze the critical properties of the models considered here. General reviews of FSS can be found in Refs.~[\onlinecite{Barber-83,Privman-90,BDT-00}]; a summary of FSS theory which focuses on symmetry-breaking boundary conditions is found in Ref.~[\onlinecite{PTD-10}], while FSS at a quantum phase transition is discussed in Ref.~[\onlinecite{CPV-14}].
According to renormalization group (RG) theory~\cite{Wegner-76,PV-02}, the free energy {\it per volume $L^d$}, ${\cal F}(g,h,h_0,T,L)$ splits into a sum of a {\it n}on{\it s}ingular and a {\it s}ingular contribution,
\begin{equation}
\begin{split}
{\cal F}(g,h,h_0,T,L)=&f^{(ns)}(g,h,h_0,T,L)\\
&+f^{(s)}(g,h,h_0,T,L).
\end{split}
\label{fsns}
\end{equation}
Since the presence of the pinning field gives rise to a line defect, one expects that the regular part of the free energy is a sum of a ``bulk'' free energy density (independent of $h_0$) and of a ``line'' free energy density:
\begin{equation}
\begin{split}
f^{(ns)}(g,h,h_0,T,L)=&f_{\rm bulk}^{(ns)}(g,h) \\
&+ \frac{1}{L^{D-1}}f_{\rm line}^{(ns)}(g,h,h_0),
\end{split}
\label{fnonsingH}
\end{equation}
where, in line with the quantum-to-classical mapping, both regular terms are expected to be temperature-independent~\cite{CPV-14}.

In the FSS limit, the singular part of the free energy density obeys the scaling ansatz~\cite{Barber-83,Privman-90,BDT-00,CPV-14}
\begin{equation}
\begin{split}
&f^{(s)}(g,h,h_0,T,L) =\frac{1}{(L+c)^D}\\
&\cdot f(u(L+c)^{1/\nu}, h(L+c)^{y_h}, h_0(L+c)^{y'}, \{u_i(L+c)^{-\omega_i}\}, \rho),\\
\end{split}
\label{fsingH}
\end{equation}
where
\begin{equation}
\begin{split}
u &= \frac{g-g_c}{g_c},\\
\rho &= \frac{L}{c_0 (1/T)},\\
y_h &= \frac{D+2-\eta}{2}.
\end{split}
\label{scaling_fieldsH}
\end{equation}
In Eqs.~(\ref{fsingH})-(\ref{scaling_fieldsH}) we have introduced the generalized aspect ratio $\rho$; up to a nonuniversal constant $c_0$ associated with the scaling field of the temperature $T$, $\rho$ is the ratio of the spatial size $L$ of the model and the size $\sim 1/T$ in the imaginary-time direction, which emerges after the quantum-to-classical mapping. In Eq.~(\ref{fsingH}), we have assumed $h_0$ to be a perturbation around $h_0=0$, and anticipated the discussion of Sec.~\ref{sec:fss:h0}, where we show that it is a scaling field with RG dimension $y'$. The irrelevant scaling fields $\{u_i\}$ give rise to corrections to scaling which decay as $\propto L^{-\omega_i}$, the dominant one being the one with the smallest $\omega$. In Eq.~(\ref{fsingH}) we have also included the correction to scaling arising from the lack of full translational invariance. Under this condition, $L$ is not an exact scaling field, but it has to be replaced with an expansion $L+c+O(1/L)$, which results in corrections to scaling $\propto L^{-1}$; $c$ is a nonuniversal constant. The property that such corrections to scaling can be adsorbed by the substitution $L\rightarrow L+c$ was first proposed in the context of surface susceptibilities~\cite{CF-76} and more recently verified for improved classical models in a film geometry, where it gives rise to the leading scaling correction~\cite{Hasenbusch-08,Hasenbusch-10c,PTD-10,Hasenbusch-11,Hasenbusch-12,PTTD-13,PTTD-14,Hasenbusch-14,PT-13}. The expansion $L+c+O(1/L)$ for the scaling field associated with the system size has also been argued to hold for a quantum phase transition at $T=0$~\cite{CPV-14}. Scaling corrections arising from nonlinearities in the scaling fields~\cite{AF-83} have been neglected in Eqs.~(\ref{fsingH})-(\ref{scaling_fieldsH}) because they do not play a relevant role here.

The FSS behavior of the classical Blume-Capel model defined in Sec.~\ref{sec:models:bc} is essentially identical to the one for the bilayer Heisenberg model, requiring only a minor change in the definitions of coupling constants and scaling fields. Similar to Eq.~(\ref{fnonsingH}), the nonsingular part of the free energy density $f^{(ns)}(K,h,h_0,L_z,L)$, i.e., the free energy per volume $L^D$, decomposes into a bulk and a line term, which are now functions of the coupling constant $K$ of the Hamiltonian (\ref{HBC})\footnote{For sake of simplicity, we omit here the dependency on the Blume-Capel parameter $\Delta$, since it has been fixed to $\Delta=0.655$ throughout the investigations, see Sec.~\ref{sec:models:bc}.}:
\begin{equation}
f^{(ns)}(K,h,h_0,L_z,L)=f_{\rm bulk}^{(ns)}(K,h) + \frac{1}{L^{D-1}}f_{\rm line}^{(ns)}(K,h,h_0).
\label{fnonsingBC}
\end{equation}
The singular part of the free-energy density $f^{(s)}(K,h,h_0,L_z,L)$ satisfies a scaling behavior analogous to Eq.~(\ref{fsingH}),
\begin{equation}
\begin{split}
&f^{(s)}(K,h,h_0,L_z,L) =\frac{1}{(L+c)^D}\\
&\cdot f(u(L+c)^{1/\nu}, h(L+c)^{y_h}, h_0(L+c)^{y'}, \{u_i(L+c)^{-\omega_i}\}, \rho),\\
\end{split}
\label{fsingBC}
\end{equation}
where $y_h$ is given as in Eq.~(\ref{scaling_fieldsH}) and the scaling field $u$ and the aspect ratio $\rho$ are
\begin{equation}
u = \frac{K_c-K}{K},\qquad \rho = \frac{L}{L_z}.
\label{scaling_fieldsBC}
\end{equation}
Since in a finite size $L$ there are no singularities in the free energy, the FSS scaling functions of Eqs.~(\ref{fsingH}) and (\ref{fsingBC}) are expected to be smooth functions in their variables, and in particular analytical in the aspect ratio $\rho$~\footnote{For some limiting values of $\rho$, FSS functions can exhibit additional nonanalyticities associated with a phase transition in reduced dimensionality, such as in a three-dimensional film geometry of infinite lateral extent at the onset of the two-dimensional phase transition~\cite{Hasenbusch-08,PTTD-13,Hasenbusch-14}.}. Therefore, as long as the FSS limit is taken at fixed $\rho$, the presence of this scaling variable does not affect the determination of the exponents from a FSS analysis. Nevertheless, we mention that the actual dependence of the free energy on $\rho$ in the presence of isolated line defects is of particular interest in the film geometry, since an observed linear behavior in $\rho$ for $\rho\rightarrow 0$ can be used to extract a contribution to FSS functions which is solely due to the line defects~\cite{PTD-10,PTTD-13,PTTD-14}. The simulation results for the bilayer Heisenberg model presented in Sec.~\ref{sec:results:heisenberg} have been obtained by
finite-temperature QMC simulations~\cite{Sandvik99,Syljuasen02,Alet05} that target the ground-state from employing an  aspect ratio of $\rho=0.5/c_0$ [see Eq.~(\ref{scaling_fieldsH})], whereas the Monte Carlo (MC) results of the Blume-Capel model reported in Sec.~\ref{sec:models:bc} correspond to a cubic lattice of equal linear size in all directions, i.e., with $\rho=1$ [see Eq.~(\ref{scaling_fieldsBC})].

From Eqs.~(\ref{fsns})--(\ref{scaling_fieldsBC}), the FSS behavior of the various observables can be obtained by taking the appropriate derivatives. The presence of a nonzero pinning field breaks explicitly the SU(2) symmetry, giving rise to a nonzero magnetization per volume $m$. Its FSS behavior is readily obtained by differentiating the free energy per volume with respect to $h$ and setting $h=0$. Notice that, because of the O(3) symmetry of the bilayer Heisenberg model $f_{\rm bulk}^{(ns)}(g,h,T)=f_{\rm bulk}^{(ns)}(g,-h,T)$ and $f_{\rm line}^{(ns)}(g,h,h_0,T)=f_{\rm line}^{(ns)}(g,-h,-h_0,T)$. Therefore, since, by definition, $f^{(ns)}$ is not singular, we have $\partial f_{\rm bulk}^{(ns)}(g,h,T)/\partial h=0$ for $h=0$, whereas in general $\partial f_{\rm line}^{(ns)}(g,h,h_0,T)/\partial h\ne0$ for $h=0$ but $h_0\ne 0$. Analogous results follow for the Blume-Capel model, which is $\mathbb Z_2$-symmetric in the absence of the pinning field.
Using Eq.~(\ref{fnonsingH}), for the bilayer Heisenberg model, and Eq.~(\ref{fnonsingBC}), for the Blume-Capel mode, we obtain the FSS of the magnetization per volume $m$ in zero external field $h$ and nonzero pinning field $h_0$:
\begin{align}
m&(g,h_0,T,L)=(L+c)^{-\beta/\nu}\nonumber \\
&\cdot f_m(u(L+c)^{1/\nu},h_0(L+c)^{y'}, \{u_i(L+c)^{-\omega_i}\}, \rho) \nonumber \\
&+ \frac{1}{L^{D-1}}g_{\rm line}(g,h_0), \quad \text{Heisenberg}
\label{mH}\\
m&(K,h_0,L_z,L)=(L+c)^{-\beta/\nu}\nonumber \\
&\cdot f_m(u(L+c)^{1/\nu},h_0(L+c)^{y'}, \{u_i(L+c)^{-\omega_i}\}, \rho) \nonumber \\
&+ \frac{1}{L^{D-1}}g_{\rm line}(K,h_0), \quad \text{Blume-Capel},
\label{mBC}
\end{align}
where we have introduced the scaling functions $f_m$ for the singular part of the magnetization, their nonsingular counterpart $g_{\rm line}$, and have used $D-y_h=\beta/\nu$. A comparison of Eq.~(\ref{mH}) with Eq.~(\ref{mBC}) shows that the scaling behavior is essentially identical for both models considered here.
By setting $g=g_c$ for the bilayer Heisenberg model (respectively, $K=K_c$ for the Blume-Capel model), and retaining the leading correction to scaling only, we obtain the FSS behavior of the magnetization at criticality:
\begin{equation}
\begin{split}
m=L^{-\beta/\nu}\Big[&f_{mc}(h_0L^{y'},\rho)+L^{-\omega}g_{mc}(h_0L^{y'},\rho)\\
&+ L^{-(D-1-\beta/\nu)}g_{\rm line,c}(h_0)\Big],\\
&\omega=\text{min}\{1,\{\omega_i\}\},\quad \text{criticality},
\end{split}
\label{mcrit}
\end{equation}
where, in order to employ a uniform notation for both models considered here, we have introduced two scaling functions $f_{mc}(h_0L^{y'},\rho)$ and $g_{mc}(h_0L^{y'},\rho)$ describing the leading, and the dominant correction-to-scaling terms. The last term in Eq.~(\ref{mcrit}) is defined as $g_{\rm line,c}(h_0)\equiv g_{\rm line}(g_c,h_0)$ for the bilayer Heisenberg model and $g_{\rm line,c}(h_0)\equiv g_{\rm line}(K_c,h_0)$ for the Blume-Capel model. It represents the scaling corrections due to the analytical background, which are characterized by an effective correction-to-scaling exponent $D-1-\beta/\nu$. In $D=3$, for both models considered here we have $D-1-\beta/\nu\approx 1.5$, therefore such corrections are negligible with respect to those arising from $g_{mc}(h_0L^{y'},\rho)$. In the following we shall neglect the corrections due to the nonsingular part of the free energy.

\subsection{RG-flow of $h_0$}
\label{sec:fss:h0}
In the preceding section we have implicitly assumed that $h_0$ represents a line scaling field at the fixed point $h_0=0$. Indeed, for $h_0=0$ the system is translationally invariant, and the RG flow does not generate line couplings which would break this symmetry, thus $h_0=0$ is a fixed point of the RG flow. The scaling dimension of $h_0$ at the $h_0=0$ fixed point can be determined as follows. In the continuum limit, a field-theory description of the pinning field corresponds to an interaction term in the action as
\begin{equation}
h_0\int d^D\vec{x} \delta^{(D-1)}(\vec{x}_\perp)\phi(\vec{x}),
\label{h0interaction}
\end{equation}
where $\phi(\vec{x})$ is the order parameter and $\delta^{(D-1)}(\vec{x}_\perp)=\delta(x)\delta(y)$ indicates that the interaction is restricted to the fields with coordinates $x=y=0$. Under an RG transformation with scale factor $b$, $\vec{x}\rightarrow \vec{x}' = \vec{x} / b$, $\phi(\vec{x})\rightarrow \phi'(\vec{x}') = b^{x_\phi}\phi(\vec{x})$, $h_0\rightarrow h_0' = h_0b^{y'}$, where $x_\phi=D-y_h=\beta/\nu$ is the scaling dimension of the operator $\phi$ and $y'$ the RG dimension of the pinning field, to be determined. Correspondingly, the interaction of Eq.~(\ref{h0interaction}) transforms to
\begin{equation*}
\begin{split}
&h_0'\int d^D\vec{x}'\delta^{(D-1)}(\vec{x}'_\perp)\phi'(\vec{x}')\\
&=b^{y'-D+(D-1)+x_\phi}h_0\int d^D\vec{x} \delta^{(D-1)}(\vec{x}_\perp)\phi(\vec{x})
\end{split}
\end{equation*}
Scale invariance requires then
\begin{equation}
y'=1 - x_\phi=y_h-(D-1)=\frac{1-\eta}{2}.
\label{yprime}
\end{equation}
An equivalent argument leading to Eq.~(\ref{yprime}) follows from the scaling behavior of the two-point function and it is reported in Appendix.~\ref{sec:chiloc}.

The calculation outlined above crucially depends on the fact that the order-parameter operator $\phi$ on the line scales with the usual RG dimension $x_\phi=\beta/\nu$. Indeed, for $h_0=0$ the model is translationally invariant, therefore the local operator $\phi$ on the line is no different than the bulk one. Thus, as a perturbation around the $h_0=0$ fixed point, the transformation under the RG of the interaction of Eq.~(\ref{h0interaction}) can be deduced from the usual RG transformation of the order-parameter operator $\phi$, as illustrated above. Analogous calculations allow to deduce the scaling dimension of any other line perturbations, at the $h_0=0$ fixed point. In contrast, an equivalent argument in the presence of a real surface or edge does not lead to the correct scaling dimension of edge or surface fields, because the boundary operators have different scaling dimensions than the bulk ones. In terms of the corresponding field theory, surface and edge terms require an additional renormalization~\cite{Diehl-86,Diehl-97}.

\begin{figure}
\includegraphics[width=\columnwidth,keepaspectratio]{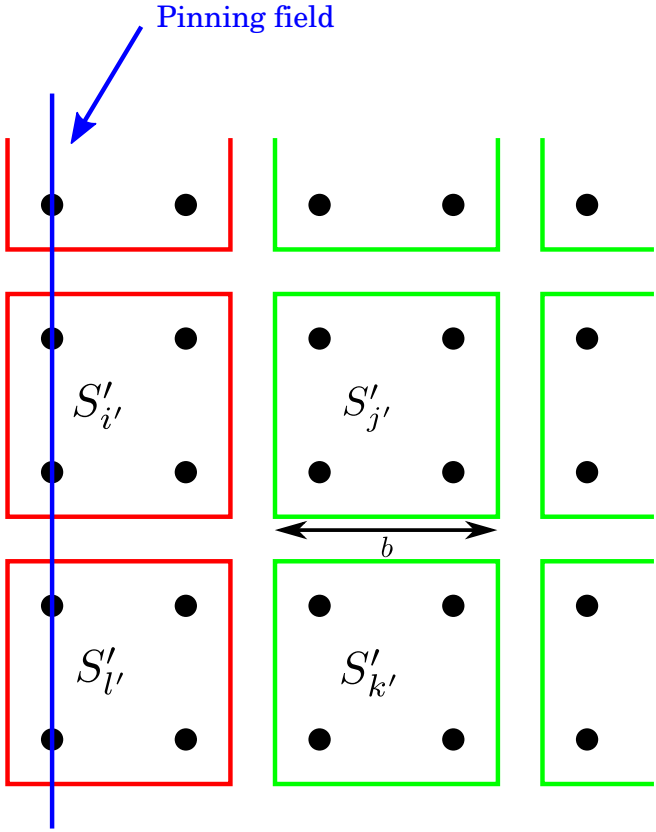}
\caption{Real-space RG transformation with scale $b$, in the presence of a defect line. Black points represent the spin degrees of freedom, $S'_{i'}$ is the coarse-grained spin at the block $i'$. With a local Hamiltonian, coarse graining and rescaling for the spins in the green blocks is independent of the presence of the defect line, which influences the RG transformation near the defect. After an RG step, the effective size of the ``bulk'' part is no longer precisely defined, this implies that the length $L$ is no longer an exact scaling field, but has to be replaced by $L+c+O(1/L)$.}
\label{fig:rg}
\end{figure}
Using Eq.~(\ref{yprime}), we find that for the Heisenberg UC $y'=0.4813(3)$~\cite{CHPRV-02}, for the Ising UC $y'=0.48187(5)$~\cite{Hasenbusch-10}. Since $y'>0$, the $h_0=0$ fixed point is unstable against the inclusion of a pinning field. The corresponding RG flow shares important analogies with surface critical phenomena~\cite{Binder-83,Diehl-86,Diehl-97}. There one considers ``bulk'' couplings, whose RG flow is independent on boundary conditions, and ``surface'' couplings, whose transformation under the RG depends also on the bulk ones. Here, in analogy with surface critical behavior, one considers the RG flow of bulk couplings, which is independent of the presence of a pinning field, and the RG flow of ``line'' couplings, which instead depends also on the bulk properties. This picture can be intuitively understood by considering a real-space RG transformation of the lattice model~\cite{Cardy-book}. Since the RG is a local transformation with a local Hamiltonian, coarse graining and rescaling of the degrees of freedom is independent of the presence of a line defect, when the distance of such degrees of freedom from the line defect exceeds the scale of the RG transformation. Conversely, the RG flow close to the line defect is influenced by the presence of a pinning field. Fig.~\ref{fig:rg} illustrates this argument. This picture provides also an intuitive explanation of the necessity of replacing the scaling field $L$ with an expansion of the form $L+c+O(1/L)$ when the translational symmetry is broken (see the discussion in Sec.~\ref{sec:fss:general}).

For a nonzero pinning field $h_0$, the line defect is always ordered, therefore, analogous to the case of a real surface with a surface field, under the RG $h_0$ flows to $h_0=\infty$, corresponding to the so-called ``normal'' UC, or ``critical adsorption'' fixed point~\cite{Binder-83,Diehl-86,Diehl-97}. This picture has been confirmed for the Ising UC by means of Migdal-Kadanoff calculations in Ref.~[\onlinecite{Hanke-00}], where it has been shown that critical adsorption on a line defect represents a new line UC. Thus, the FSS behavior of the models considered here can be explained in terms of a crossover from the ``ordinary'' fixed point $h_0=0$ to the critical adsorption fixed point $h_0=\infty$.

At this fixed point, the singular part of the free energy depends only on the bulk couplings, whose RG flow, as discussed above, is independent of the line couplings. Neglecting for simplicity scaling corrections, under a RG transformation with scale $b$ the singular part of the free energy density for the model of Eq.~(\ref{HHeisenberg}) transforms as
\begin{equation}
f^{(s)}(g,h,T,L) = b^{-D}\widetilde{f}(ub^{1/\nu}, hb^{y_h},Tb^z, L/b).
\end{equation}
Fixing the scale of the previous equation with $L/b={\rm const}$, one finds that $f^{(s)}$ satisfies the FSS form
\begin{equation}
f^{(s)}(g,h,T,L) = \frac{1}{L^D}f(uL^{1/\nu}, hL^{y_h},\rho),
\label{fsingH_normal}
\end{equation}
with $\rho$ defined in Eq.~(\ref{scaling_fieldsH}). An analogous result holds for the Blume-Capel model. By differentiating Eq.~(\ref{fsingH_normal}) with respect to $h$, we obtain for both models the following prediction for the asymptotic behavior of the magnetization $m$:
\begin{equation}
m = L^{-\beta/\nu}f_{m,norm}(uL^{1/\nu},\rho), \qquad h_0\rightarrow\infty
\label{m_norm}
\end{equation}
where scaling corrections have been neglected. A comparison with Eqs.~(\ref{mH})-(\ref{mBC}) reveals that
\begin{equation}
f_{m,norm}(\widetilde{u}, \rho) = \lim_{\widetilde{h_0}\rightarrow \infty} f_m(\widetilde{u}, \widetilde{h_0}, \{0\}, \rho),
\label{m_norm_asymptotic}
\end{equation}
where $\widetilde{u}=uL^{1/\nu}$, $\widetilde{h_0}=h_0L^{y'}$ are the scaling variables of $f_m$ and $f_{m,norm}$. At criticality $u=0$ and Eq.~(\ref{m_norm_asymptotic}) implies that the scaling function $f_{mc}$ introduced in Eq.~(\ref{mcrit}) is asymptotically flat,
\begin{equation}
\lim_{\widetilde{h_0}\rightarrow \infty}f_{mc}(\widetilde{h_0},\rho) = \text{const}(\rho),
\label{mcrit_norm_asymptotic}
\end{equation}
i.e., independent of the first variable (but, in principle, still dependent on the aspect ratio $\rho$).

\subsection{Order-parameter profiles and local susceptibilities}
\label{sec:fss:op}
The analysis of spatially-resolved observables, such as the magnetization profiles, provides valuable additional information on the critical behavior of a system. Here we do not address the crossover behavior induced by a finite pinning field $h_0$ but limit our analysis to the scaling behavior at the critical adsorption fixed point, i.e., for $h_0=\infty$.
In the following we refer for simplicity to the FSS behavior of the Blume-Capel model defined in Eq.~(\ref{HBC}), for which we can also make contact with the field-theory results of Ref.~[\onlinecite{Hanke-00}]; analogous results hold for the scaling behavior of the bilayer Heisenberg model.

The local magnetization at lattice site $\vec{x}$ is given by the expectation value of the order parameter $\phi(\vec{x})$. Due to the translational invariance along the pinning-field line, and to the rotational invariance restored at criticality, the magnetization profile $m(x_\perp)$ depends only on the distance from the pinning field $x_\perp=|\vec{x}_\perp|$ (see the discussion after Eq.~(\ref{HBC}) for the definition of the coordinate system). At zero bulk field, under an RG transformation with scale $b$, $m(x_\perp)$ transforms as
\begin{equation}
m(x_\perp) = b^{-x_\phi}\widetilde{m}(ub^{1/\nu},\{u_ib^{-\omega_i}\},x_\perp/b,L/b,L_z/b),
\label{profRG}
\end{equation}
where we have neglected the analytical scaling corrections $\propto L^{-1}$ arising from non-translational invariance [see the discussion after Eq.~(\ref{scaling_fieldsH})]. Fixing the scale $L/b={\rm const}$ in the previous equation, and expanding for large $L$, retaining the leading irrelevant operator, we find
\begin{equation}
\begin{split}
m(x_\perp) = L^{-\beta/\nu}&f_p(uL^{1/\nu},x_\perp/L,\rho)\cdot \\
& \left[1+L^{-\omega}g_p(uL^{1/\nu},x_\perp/L,\rho)\right],
\end{split}
\label{prof}
\end{equation}
where for later convenience we have factorized the subleading term $g_p$.
At criticality, $u=0$ and Eq.~(\ref{prof}) reduces to
\begin{equation}
m(x_\perp) = L^{-\beta/\nu}f_{pc}(x_\perp/L,\rho)\left[1 + L^{-\omega}g_{pc}(x_\perp/L,\rho)\right],
\label{profcrit}
\end{equation}
where $f_{pc}(\widetilde{x}_\perp,\rho)\equiv f_{p}(\widetilde{u}=0,\widetilde{x}_\perp,\rho)$, $g_{pc}(\widetilde{x}_\perp,\rho)\equiv g_{p}(\widetilde{u}=0,\widetilde{x}_\perp,\rho)$.
Equations (\ref{prof}) and (\ref{profcrit}) hold in the FSS limit, i.e., in the limit $L\rightarrow\infty$, with fixed $uL^{1/\nu}$, $\rho$, and, additionally, fixed $x_\perp/L$. A scaling argument can be formulated to show that, in fact, the tail of the profiles satisfy Eq.~(\ref{profcrit}) even for a finite value of $h_0$, independent of that~\cite{BE-85}.
It is useful to also consider the infinite-volume limit, i.e., $L$, $L_z\rightarrow\infty$. In this limit, neglecting analytical corrections to scaling with $\omega=1$, under a RG transformation with scale $b$, $m(x_\perp)$ transform as [compare with Eq.~(\ref{profRG})]
\begin{equation}
m(x_\perp) = b^{-x_\phi}\widetilde{m}(ub^{1/\nu},\{u_ib^{-\omega_i}\},x_\perp/b),
\label{profRG_infL}
\end{equation}
Fixing now the scale with $x_\perp/b={\rm const}$ and expanding for the tail of the profile, we find the infinite-volume counterpart of Eqs.~(\ref{prof}) and (\ref{profcrit}):
\begin{equation}
m(x_\perp) = x_\perp^{-\beta/\nu}f_{p,\infty}(ux_\perp^{1/\nu})\left(1 + x_\perp^{-\omega}g_{p,\infty}(ux_\perp^{1/\nu})\right),
\label{prof_infL}
\end{equation}
\begin{equation}
m(x_\perp) = x_\perp^{-\beta/\nu}f_{pc,\infty}\left(1 + x_\perp^{-\omega}g_{pc,\infty}\right), \quad \text{criticality},
\label{profcrit_infL}
\end{equation}
where $f_{pc,\infty}=f_{p,\infty}(\widetilde{u}=0)$, $g_{pc,\infty}=g_{p,\infty}(\widetilde{u}=0)$, so that at a critical adsorption fixed point the magnetization profile decays with the $\beta/\nu$ exponent~\cite{Diehl-86,Diehl-97}. We recall that Eqs.~(\ref{profRG})--(\ref{profcrit_infL}) holds only for distances $x_\perp\gg \sigma_0$, with $\sigma_0$ the short-distance scale which controls the nonuniversal behavior.
A comparison of Eq.~(\ref{prof_infL}) with Eq.~(\ref{prof}), and of Eq.~(\ref{profcrit_infL}) with Eq.~(\ref{profcrit}) gives the following boundary conditions for the FSS functions $f_{p}(\widetilde{u},\widetilde{x}_\perp,\rho)$, $f_{pc}(\widetilde{x}_\perp,\rho)$, $g_{p}(\widetilde{u},\widetilde{x},_\perp\rho)$, $g_{pc}(\widetilde{x}_\perp,\rho)$:
\begin{align}
\label{profBC}
&f_{p}(\widetilde{u}\rightarrow\infty,\widetilde{x}_\perp\rightarrow 0,\rho)\simeq \widetilde{x}_\perp^{-\beta/\nu}f_{p,\infty}(ux^{1/\nu}),\\
\label{profBC_corr}
&g_{p}(\widetilde{u}\rightarrow\infty,\widetilde{x}_\perp\rightarrow 0,\rho)\simeq \widetilde{x}_\perp^{-\beta/\nu}g_{p,\infty}(ux^{1/\nu}),\\
\label{profcritBC}
&f_{pc}(\widetilde{x}_\perp\rightarrow 0,\rho)\simeq \widetilde{x}_\perp^{-\beta/\nu}f_{pc,\infty},\\
\label{profcritBC_corr}
&g_{pc}(\widetilde{x}_\perp\rightarrow 0,\rho)\simeq \widetilde{x}_\perp^{-\beta/\nu}g_{pc,\infty},
\end{align}
where in Eqs.~(\ref{profBC_corr}) and (\ref{profcritBC}) the limit $\widetilde{u} \rightarrow \infty$, $\widetilde{x}_\perp \rightarrow 0$ is taken at fixed $\widetilde{u}\widetilde{x}^{1/\nu} = ux^{1/\nu} \neq 0$.
By inserting Eq.~(\ref{profcritBC}) into Eq.~(\ref{profcrit}) we find the important result that the {\it leading} short-distance behavior of the order-parameter profile does not depend on the finiteness of the system, which, however, gives rise to the analog of the so-called distant-wall corrections~\cite{FG-78}.

The behavior of the order-parameter profile close to the defect line can be analyzed within the framework of the short-distance expansion (SDE), which is an analog of the operator product expansion (OPE)~\cite{DD-81,*DD-81E}. According to SDE, close to the line defect (or, more generically, to a surface) the order-parameter field $\phi(\vec{x})$ can be expanded as
\begin{equation}
\phi(\vec{x}) \underset{x_\perp\rightarrow 0}{\mathop{=}} \sum_i C_{\psi_i} x_\perp^{-x_\phi+x_{\psi_i}}\psi_i(\vec{x}_\parallel).
\label{SDE}
\end{equation}
In Eq.~(\ref{SDE}), the right-hand side represents a sum over ``line'' operators $\psi_i$, whose scaling dimension is $x_{\psi_i}$; in general, such operators have scaling dimensions different than the bulk ones, and they depend on the fixed point, or UC, of the line. The leading contribution in Eq.~(\ref{SDE}) is given by the identity operator, and by the operator $\psi_i\equiv O$, which has the smallest scaling dimension $x_O\equiv x_{\psi_i}$, so that
\begin{equation}
\phi(\vec{x}) \underset{x_\perp\rightarrow 0}{\mathop{=}} A x_\perp^{-\beta/\nu}\left(1 + C_Ox_\perp^{x_O}O(\vec{x}_\parallel)+\cdots\right),
\label{SDE_approx}
\end{equation}
where we have used $x_\phi=\beta/\nu$ and for convenience we have factorized a constant in front of the expansion. By taking the expectation value of Eq.~(\ref{SDE_approx}) at criticality in a finite size $L$, we find the short-distance behavior of the order-parameter profile
\begin{equation}
m(x_\perp\rightarrow 0)=A x_\perp^{-\beta/\nu}\left(1 + C_Of_O(\rho)\left(\frac{x_\perp}{L}\right)^{x_O}+\cdots\right),
\label{profcrit_SDE}
\end{equation}
where in principle $\<O(\vec{x}_\parallel)\>$ could depend on the aspect ratio $\rho$, which we have encoded in the coefficient $f_O(\rho)$ on the right-hand side of Eq.~(\ref{profcrit_SDE}). A comparison of Eq.~(\ref{profcrit_SDE}) with Eq.~(\ref{profcrit}) allows to compute the leading correction to Eq.~(\ref{profcritBC}),
\begin{equation}
f_{pc}(\widetilde{x}_\perp\rightarrow 0,\rho)\simeq \widetilde{x}_\perp^{-\beta/\nu}f_{pc,\infty}\left[1 + C_Of_O(\rho)\widetilde{x}_\perp^{x_O}\right],
\label{profcritBC_SDE}
\end{equation}
and to infer $A=f_{pc,\infty}$.

We observe that the prediction for the short-distance behavior of the order-parameter profile given in Eq.~(\ref{profcrit_SDE}) neglects corrections to scaling, which are represented by the subleading scaling function $g_{pc}$ on the right-hand side of Eq.~(\ref{profcrit}). The control of scaling corrections is essential in order to extract the critical behavior from finite-size MC data. Crucially, the full set of order-parameter profiles as a function of $\vec{x}_\perp$ and $L$ cannot be correctly described by the right-hand side of Eq.~(\ref{profcrit_SDE}), because even for $L\rightarrow\infty$ the subleading correction term $L^{-\omega}g_{pc}$ on the right-hand side of Eq.~(\ref{profcrit}) approaches a function which is $L$-independent, but still depends on $\vec{x}_\perp$ [see Eq.~(\ref{profcritBC_corr})]. Thus, in order to determine the parameters on the right-hand side of Eq.~(\ref{profcrit_SDE}), such as the scaling dimension $x_O$, one can first extrapolate the FSS limit by a fit of the order-parameter profile data to the right-hand side of Eq.~(\ref{profcrit}) {\it at fixed $x_\perp/L$, $\rho$}. This allows to extract the leading scaling function $f_{pc}(\widetilde{x},\rho)$, whose short-distance behavior can be fitted against the right-hand side of Eq.~(\ref{profcritBC_SDE}); an example of such calculation is provided in Ref.~[\onlinecite{PTD-10}].
Alternatively, one can consider the $L$ dependence of the magnetization profile at a {\it fixed short distance} $x_\perp\ll L$ from the defect line. According to Eq.~(\ref{profcritBC_SDE}) and including the leading correction to scaling present in Eq.~(\ref{profcrit}), in this regime the order-parameter profile is given by
\begin{equation}
\begin{split}
m(x_\perp\ll L) \simeq &f_{pc,\infty}x_\perp^{-\beta/\nu}\left[1+C_Of_O(\rho)\left(\frac{x_\perp}{L}\right)^{x_O}\right]\\
&\cdot \left[1+L^{-\omega}g_{pc}(x_\perp/L,\rho)\right].
\end{split}
\label{profcrit_SDE_corr}
\end{equation}
According to Eq.~(\ref{profcritBC_corr}), in the large-$L$ limit, at {\it fixed $x_\perp\ll L$}, the correction-to-scaling term $L^{-\omega}g_{pc}$ on the right-hand side of Eq.~(\ref{profcrit_SDE_corr}) converges to a $x_\perp$-dependent, but $L$-independent term. Therefore, as a function of $L$, the magnetization profile attains the following form
\begin{equation}
\begin{split}
m(x_\perp\ll L) \simeq \text{const}\left[1+\text{const}\cdot L^{-x_O}\right],\\
\quad L\rightarrow\infty, \qquad \vec{x}_\perp \text{fixed},
\end{split}
\label{profcrit_SDE_fit}
\end{equation}
where the constants are $L$-independent.

The SDE of Eq.~(\ref{SDE}) and Eq.~(\ref{SDE_approx}) concerns also the decay of the two-point function at criticality, along the line defect
\begin{equation}
\begin{split}
\<\phi(\vec{x}_\perp,\vec{x}_\parallel)\phi(\vec{x}_\perp,\vec{x}'_\parallel)\>_c \sim &|\vec{x}_\parallel-\vec{x}'_\parallel|^{-(D-2+\eta_\parallel)}, \\
&|\vec{x}_\parallel-\vec{x}'_\parallel|\rightarrow\infty,
\end{split}
\label{corr_par}
\end{equation}
where the subscript $c$ indicates the connected part and we have introduced the exponent $\eta_\parallel$ which is related to $x_O$ by
\begin{equation}
x_O=\frac{D-2+\eta_\parallel}{2}.
\label{etapar}
\end{equation}
In a finite size $L$, the exponent $\eta_\parallel$ characterizes the size dependence of the local ``line'' susceptibility $\chi_{\rm loc}$. In the absence of bulk field $\chi_{\rm loc}(\vec{x}_\perp,K,h_0,L_z,L)$ is defined as
\begin{equation}
\begin{split}
\chi_{\rm loc}(\vec{x}_\perp,K,h_0,L_z,L) \equiv \frac{1}{L_z}\sum_{z,z'}&\<S_{(x,y,z)}S_{(x,y,z')}\> \\
&-\<S_{(x,y,z)}\>\<S_{(x,y,z')}\>.
\end{split}
\label{chilocdef}
\end{equation}
At bulk criticality, and at the critical adsorption line fixed point, the scaling part $\chi_{\rm loc}^{\rm scal.}$ of $\chi_{\rm loc}$ for $x_\perp\rightarrow 0$, as inferred from the correlations of Eq.~(\ref{chilocdef}) or directly from the the SDE of Eq.~(\ref{SDE_approx}), is given by
\begin{equation}
\chi_{\rm loc}^{\rm scal.}(x_\perp\rightarrow 0,K_c,h_0=\infty,L_z,L) \propto L^{1-2x_O}\propto L^{-\eta_\parallel},
\label{chiloc_scaling}
\end{equation}
where, like the magnetization profile, $\chi_{\rm loc}$ depends only on $x_\perp=|\vec{x}_\perp|$ in the critical region and the proportionality constant contains the dependence on $x_\perp$ and the aspect ratio $\rho$, inessential for the present discussion.
Since the exponent of $L$ in Eq.~(\ref{chiloc_scaling}) is negative, the scaling part is non-divergent, and thus the FSS behavior of $\chi_{\rm loc}$ is dominated by the short-distance part of the correlations $\chi_{\rm loc}^{\rm short}$, i.e., by the terms in Eq.~(\ref{chilocdef}) where $|z-z'| \ll L_z$:
\begin{equation}
\begin{split}
\chi_{\rm loc}^{\rm short}&(x_\perp\rightarrow 0,K_c,h_0=\infty,L_z,L) =\\
&\frac{1}{L_z}\sum_{\substack{z,z'\\|z-z'|\ll L_z}}\<S_{(x,y,z)}S_{(x,y,z')}\>-\<S_{(x,y,z)}\>\<S_{(x,y,z')}\>
\end{split}
\label{chiloc_short_def}
\end{equation}
Due to the translational invariance along the line, there are $O(L_z)$ terms in the sum on the right-hand side of Eq.~(\ref{chiloc_short_def}). Exploiting the translational invariance, and using the SDE of Eq.~(\ref{SDE_approx}), $\chi_{\rm loc}^{\rm short}$ is given by
\begin{equation}
\begin{split}
\chi_{\rm loc}^{\rm short}&(x_\perp\rightarrow 0,K_c,h_0=\infty,L_z,L) \\
&\propto \sum_{\substack{z\\ |z_0-z|\ll L_z}} \<O(z_0)O(z)\>-\<O(z_0)\>\<O(z)\>,
\end{split}
\label{chiloc_short_O}
\end{equation}
where $z_0$ is an arbitrary origin for the correlations, and we have used $\vec{x}_\parallel=z\hat{z}$ [see the definition of the coordinate system given after Eq.~(\ref{HBC})]. In Eq.~(\ref{chiloc_short_O}), there are $O(1)$ terms in the sum. At criticality, the second term in the sum of Eq.~(\ref{chiloc_short_O}) gives a contribution proportional to $L^{-2x_O}$, subleading with respect to the scaling behavior of Eq.~(\ref{chiloc_scaling}). The first term in the sum of Eq.~(\ref{chiloc_short_O}) can be analyzed using the OPE
\begin{equation}
O(z_0)O(z) \underset{z \rightarrow z_0}{\mathop{=}} \sum_i \frac{c_{OOi}}{|z-z_0|^{2x_O-x_{\psi_i}}} \psi_i\left(\frac{z+z_0}{2}\right),
\label{OPE_O}
\end{equation}
where the sum on the right-hand side of Eq.~(\ref{OPE_O}) is over a set of local line operators $\psi_i$ with scaling dimension $x_{\psi_i}$, calculated at the midpoint, and the OPE coefficients $c_{OOi}$ are universal, once the normalization of the operators involved in the OPE is fixed. In the expansion one finds in particular the identity operator, with $x_{\rm Id}=0$. Upon taking the expectation value at criticality, this gives a constant: it is in fact the background contribution to the local susceptibility which originates from the regular part of the free energy. Such a term $\propto L^0$, being not divergent, dominates over the scaling part of Eq.~(\ref{chiloc_scaling}). Moreover, there is also another relevant contribution in the OPE, given by the operator $O$ itself. Due to the fact that at the critical adsorption fixed point the symmetry is broken, on the right-hand side of Eq.~(\ref{OPE_O}) there are both even and odd operators, thus in particular the operator $O$ itself appears in the OPE of Eq.~(\ref{OPE_O}). By taking the expectation value at criticality, this gives rise to a contribution $\propto L^{-x_O}$ to $\chi_{\rm loc}^{\rm short}$. In many cases, including the present one, $x_O>1$, so that also such second contribution $\propto L^{-x_O}$ to $\chi_{\rm loc}^{\rm short}$ dominates over the scaling behavior $\propto L^{-\eta_\parallel}$ given in Eq.~(\ref{chiloc_scaling}). We conclude that at criticality, the leading FSS behavior of $\chi_{\rm loc}$ is as follows:
\begin{equation}
\chi_{\rm loc}(x_\perp\rightarrow 0,K_c,h_0=\infty,L_z,L) = A + B(\rho)L^{-x_O}.
\label{chiloc_fss}
\end{equation}

The SDE expansion of Eq.~(\ref{SDE}), and the corresponding short-distance order-parameter profile given in Eq.~(\ref{profcrit_SDE}) apply not only to the case of a pinning-field line, but more generically in the presence of defects, or confining surfaces. Moreover, besides the order-parameter operator $\phi$, SDE applies also to other operators, such as the energy operator~\cite{EKD-93,*EKD-93_erratum,ES-94}. For a three-dimensional system confined by surfaces in the normal UC, i.e., by ordered surfaces, the leading operator in the SDE of the order parameter is the limit $x_\perp\rightarrow 0$ of the perpendicular component of the stress-energy tensor $T_{\perp\perp}(x_\perp\rightarrow 0)$~\cite{Cardy-90,ES-94}. Since the scaling dimension of $T_{\perp\perp}$ is equal to its canonical dimension, the exponent $x_O$ in Eq.~(\ref{SDE_approx}) is $x_O=D=3$, in line with an early study of the decay of the correlations parallel to an ordered surface~\cite{BM-77} [see Eq.~(\ref{corr_par})]. Moreover, $\<T_{\perp\perp}\>$ is equal to the so-called critical Casimir force~\cite{FG-78,HHGDB-08}, so that, in agreement with an early conjecture pointed out in the context of critical adsorption~\cite{FG-78}, the correction to the short-distance behavior in Eq.~(\ref{profcrit_SDE}) is $-C_+(D-1)\Delta_{+a}(x_\perp/L)^3$, with $\Delta_{+a}$ the amplitude of the critical Casimir force at $T_c$, which depends on the UC of both confining surfaces, and $C_+=1.71(4)$~\cite{PTD-10} a universal coefficient which depends only on the UC of the close surface; such a correction is known as ``distant-wall correction''.

For the present case of a line defect at the normal UC, a field-theoretical calculation reported a new, non-trivial exponent $\eta_\parallel=1.77(5)$~\cite{Hanke-00}, which governs the leading term in the SDE of the magnetization [Eqs.~(\ref{SDE_approx}) and Eq.~(\ref{profcrit_SDE})] and the decay of the correlations along the defect [Eq.~(\ref{corr_par})]. The presence of such corrections is attributed to an unknown line operator $O$, whose scaling dimension is [Eq.~(\ref{etapar})] $x_O=(1+\eta_\parallel)/2=1.385(25)$. We are not aware of numerical calculation of this exponent, nor of another confirmation of this result. Being this operator $O$ a line operator, the RG-dimension of the corresponding scaling field is given by $y_O=1-x_O$, with $y_O=-0.385(25)$, according to the results of Ref.~[\onlinecite{Hanke-00}]. Therefore the scaling behavior of the model presents an irrelevant scaling field, which gives rise to scaling corrections with $\omega=-y_O\sim 0.4$. We notice that this is an unusual small value for $\omega$. As a comparison, the leading irrelevant bulk scaling field in 3D O(N) models gives $\omega\simeq 0.8$~\cite{PV-02}, while analytical scaling corrections arising from the broken translational invariance have $\omega=1$ [see the discussion after Eq.~(\ref{scaling_fieldsH})]. Thus the results of Ref.~[\onlinecite{Hanke-00}] hints at the presence of slowly decaying scaling corrections. The results presented in Sec.~\ref{sec:results:bc} support this picture.

Due to its importance in the physics of fluids, critical adsorption has attracted numerous experimental investigations; Ref.~[\onlinecite{Law-01}] provides a review of experimental results, more recent studies are found in Refs.~[\onlinecite{CL-01,JLPZ-01,CL-01b,CLG-02,CL-05,BZCMWS-05,BLSHWCM-07}]. While most theoretical investigations concern with adsorption on a plane~\cite{BE-85,Cardy-90,CD-90,DC-91,DS-93,*DS-93_erratum,Diehl-94,FD-95}, the case of a nonplanar geometry was studied in Refs.~[\onlinecite{HD-99,HKSD-99,Hanke-00,PD-04}]. In Ref.~[\onlinecite{HD-99}], critical adsorption on a sphere and on a cylinder in an infinite volume was investigated by means of field theory: for the case of a cylinder, in the limit of small radius the order parameter at criticality and in the infinite-volume limit is found to decay $\propto x_\perp^{-\beta/\nu}$, in agreement with our analysis and Eq.~(\ref{profcrit_infL}).

\section{Results}
\label{sec:results}
\subsection{Classical lattice model}
\label{sec:results:bc}
\begin{figure}
\includegraphics[width=\linewidth,keepaspectratio]{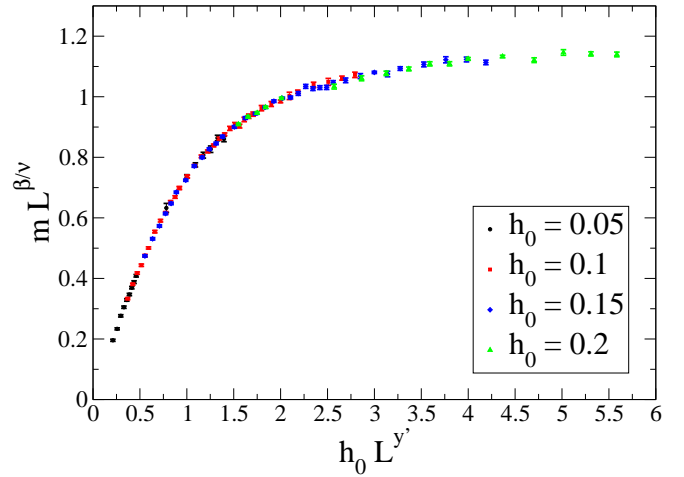}
\caption{Scaling collapse of the magnetization for the improved Blume-Capel model at the critical point $K_c=0.387721735$, $D=0.655$~\cite{Hasenbusch-10}, and for finite values of the pinning field $h_0$. The critical exponents are those of the 3D Ising UC, with $y'=(1-\eta)/2=0.48186(5)$ [see Eq.~(\ref{yprime})] and $\beta/\nu=0.51813(5)$~\cite{Hasenbusch-10}. Up to scaling corrections, $mL^{\beta/\nu}=f_{mc}(h_0L^{y'},\rho=1)$.}
\label{fig:scalingmBC}
\end{figure}

We first consider the scaling behavior of the Blume-Capel model of Eq.~(\ref{HBC}) at small value of the pinning field $h_0$. According to the discussion in Secs.~\ref{sec:fss:general} and \ref{sec:fss:h0}, for small values of $h_0$, the pinning field is expected to introduce a new scaling field with dimension given in Eq.~(\ref{yprime}). We have sampled the magnetization $m$ at the critical point by means of MC simulations, setting $K=K_c=0.387721735$, $D=0.655$~\cite{Hasenbusch-10}, and considering $h_0=0.05$, $0.1$, $0.15$, $0.2$, for a wide range of lattice sizes $L=15\ldots 1000$. According to Eq.~(\ref{mcrit}), up to scaling corrections $mL^{\beta/\nu}$ is a function of the scaling variable $h_0L^{y'}$. Using the value of the exponents of the 3D Ising UC $\eta=0.03627(10)$~\cite{Hasenbusch-10}, we obtain $y'=0.48186(5)$ and $\beta/\nu=0.51813(5)$. In Fig.~\ref{fig:scalingmBC} we plot $mL^{\beta/\nu}$ as a function of $h_0L^{y'}$, obtaining a very good collapse of the MC data, which supports Eq.~(\ref{mcrit}) with $y'$ given in Eq.~(\ref{yprime}). We notice also that, within the precision of the data, scaling corrections appear to be negligible.
In agreement with Eq.~(\ref{mcrit_norm_asymptotic}), $mL^{\beta/\nu}$ approaches a constant for large values of the scaling variable $\widetilde{h_0}=h_0L^{y'}$. In line with the discussion of Sec.~\ref{sec:fss:h0}, Fig.~\ref{fig:scalingmBC} illustrates a crossover between the $h_0=0$ and the $h_0=\infty$ fixed points.

In order to study the FSS behavior at the normal fixed point, we have simulated the Blume-Capel model of Eq.~(\ref{HBC}) at the critical point for $h_0=\infty$, and for lattice sizes $L=8,\ldots, 600$. Since, as shown below, the finite-size corrections play an important role, we have sampled the smallest lattices $L\le 48$ with a high precision, reaching a relative uncertainty of the magnetization data of $\approx 10^{-4}$.
In our simulations, for a lattice size $L$ each MC step consists in a Metropolis sweep on the entire lattice and $2L$ Wolff single-cluster moves, implemented as described in App. B of Ref.~[\onlinecite{PTD-10}]. For this set of simulations, the total number of MC steps is $20\cdot 10^6$ for $L=8$, $12$, $16$, $24$, $80\cdot 10^6$ for $L=32$,  $131\cdot 10^6$ for $L=48$, $750\cdot 10^3$ for $L=100$, and $50\cdot 10^3$ for $L=150$, $200$, $250$, $300$, $350$, $400$, $450$, $500$, $550$, $600$. To ensure thermalization we have generically discarded $20\%$ of the MC measures.

\begin{table}
\caption{Fits of the magnetization $m$ at the critical adsorption fixed point $h_0=\infty$ to Eq.~(\ref{fitm}), for $L=8\ldots 600$. $L_{\rm min}$ indicates the minimum lattice size $L$ taken into account. For the three-dimensional Ising UC $\beta/\nu=0.5181(5)$~\cite{Hasenbusch-10}.}
\begin{ruledtabular}
\begin{tabular}{l..}
\multicolumn{1}{c}{$L_{\rm min}$} & \multicolumn{1}{c}{$\beta/\nu$} & \multicolumn{1}{c}{$\chi^2/{\rm DOF}$} \\
\hline
$8$  & 0.49687(9) & 109.1 \\
$12$ & 0.4994(1) & 42.3 \\
$16$ & 0.5015(2) & 22.7 \\
$24$ & 0.5037(2) & 9.6 \\
$32$ & 0.5050(3) & 6.8 \\
$48$ & 0.5102(7) & 1.3 \\
$100$ & 0.5136(15) & 0.6 \\
$150$ & 0.513(4) & 0.6 \\
$200$ & 0.510(5) & 0.6 \\
$250$ & 0.509(6) & 0.7 \\
$300$ & 0.509(8) & 0.8 \\
$350$ & 0.51(1)  & 1.0 \\
$400$ & 0.519(15) & 1.0 \\
\end{tabular}
\end{ruledtabular}
\label{table.fitm}
\end{table}

Upon setting $K=K_c$, Eq.~(\ref{m_norm}) reduces to
\begin{equation}
m=AL^{-\beta/\nu}.
\label{fitm}
\end{equation}
We have fitted the magnetization $m$ against the right-hand side of Eq.~(\ref{fitm}), leaving $A$ and $\beta/\nu$ as free parameters. In Table \ref{table.fitm} we report the fit results as a function of the minimum lattice size $L_{\rm min}$ taken into account. The fitted value of $\beta/\nu$ shows a small but significant deviation from the expected $\beta/\nu=0.5181(5)$~\cite{Hasenbusch-10} value, with a large $\chi^2/{\rm DOF}$ (${\rm DOF}$ denotes the degrees of freedom). Moreover, the results exhibit a systematic drift towards the expected value. Only for $L\gtrsim 150$ the fitted $\beta/\nu$ agree, within $1\sim 1.5$ error bars to the value for the Ising UC. Notice also that the fits for $L_{\rm min}\le 48$ give results much more precise than those for $L_{\rm min}>48$, and, correspondly, a significantly larger value of $\chi^2/{\rm DOF}$. This is also due to the relative higher precision of the MC data for $L\le 48$.

\begin{table}
\caption{Fit of the magnetization $m$ at the critical adsorption fixed point $h_0=\infty$ to Eq.~(\ref{fitm_omega}) for $L=8\ldots 600$. $L_{\rm min}$ indicates the minimum lattice size $L$ taken into account. In the first set of fits (above), $\beta/\nu$ is a free parameter. In the second set (below), $\beta/\nu$ is fixed to the Ising UC value $\beta/\nu=0.5181(5)$~\cite{Hasenbusch-10}. The variation of the fitted $\omega$ due to the uncertainty of $\beta/\nu$ is one order of magnitude smaller than the statistical error bar of the fits.}
\begin{ruledtabular}
\begin{tabular}{l@{}.@{}.@{}.}
\multicolumn{1}{c}{$L_{\rm min}$} & \multicolumn{1}{c}{$\beta/\nu$} & \multicolumn{1}{c}{$\omega$} & \multicolumn{1}{c}{$\chi^2/{\rm DOF}$} \\
\hline
$8$  & 0.518(2) & 0.59(8) & 0.9 \\
$12$ & 0.519(4) & 0.53(11) & 1.0 \\
$16$ & 0.517(3) & 0.7(2) & 1.0 \\
$24$ & 0.515(3) & 0.9(4) & 1.0 \\
\hline
$8$ & & 0.59(2) & 0.9 \\
$12$ & & 0.58(3) & 0.9 \\
$16$ & & 0.61(4) & 0.9 \\
$24$ & & 0.62(8) & 1.0 \\
$32$ & & 0.57(10) & 1.0 \\
\end{tabular}
\end{ruledtabular}
\label{table.fitm_omega}
\end{table}
The results of Table \ref{table.fitm} clearly hint to the presence of slowly decaying scaling corrections at the $h_0=\infty$ fixed point. In the presence of corrections to scaling with a leading exponent $\omega$, Eq.~(\ref{fitm}) is modified as follows
\begin{equation}
m = A L^{-\beta/\nu}\left(1+BL^{-\omega}\right).
\label{fitm_omega}
\end{equation}
We have fitted the magnetization $m$ against the right-hand side of Eq.~(\ref{fitm}), leaving $A$, $B$, $\beta/\nu$, and $\omega$ as free parameters. In Table \ref{table.fitm_omega} we report the results of the fits. For every minimum lattice size $L_{\rm min}$ considered $\chi^2/{\rm DOF}$ is small and the results give a stable value of $\beta/\nu$, in nice agreement with the value of the Ising UC $\beta/\nu=0.5181(5)$~\cite{Hasenbusch-10}. The fits provide a significantly small value of $\omega\approx 0.5-0.6$, in line with the observation of the presence of slowly decaying scaling corrections. In order to obtain a more precise estimate of $\omega$, we have repeated the fits of $m$ to Eq.~(\ref{fitm}) fixing $\beta/\nu$ to the value of the Ising UC $\beta/\nu=0.5181(5)$~\cite{Hasenbusch-10}. The corresponding results are reported in Table \ref{table.fitm_omega}. The fitted value of $\omega$ is rather stable, and the $\chi^2/{\rm DOF}$ is small. A conservative estimate of $\omega$, compatible with the results for $L_{\rm min}\le 16$ including one error bar, is
\begin{equation}
\omega=0.60(5).
\label{omega}
\end{equation}

\begin{table}
\caption{Fits of the local magnetization $m(x_\perp)$ [respectively, the local susceptibility $\chi_{\rm loc}(x_\perp)$] at the critical adsorption fixed point $h_0=\infty$ to Eq.~(\ref{fitxO_m}) [respectively, Eq.~(\ref{fitxO_chiloc})], for a fixed distance $x_\perp$ from the pinning-field line, and $L=8\ldots 600$. $L_{\rm min}$ indicates the minimum lattice size $L$ taken into account.}
\begin{ruledtabular}
\begin{tabular}{ll@{}.@{}.}
\multicolumn{1}{c}{} & \multicolumn{1}{c}{$L_{\rm min}$} & \multicolumn{1}{c}{$x_O$} & \multicolumn{1}{c}{$\chi^2/{\rm DOF}$} \\
\hline
              & $8$ & 1.59(1) & 1.2 \\
              & $12$ & 1.55(2) & 0.7 \\
$m(x_\perp=2)$ & $16$ & 1.52(3) & 0.6 \\
              & $24$ & 1.52(6) & 0.7 \\
              & $32$ & 1.58(8) & 0.7 \\
\hline
              & $8$ & 1.735(8) & 5.7 \\
              & $12$ & 1.63(2) & 1.0 \\
$m(x_\perp=3)$ & $16$ & 1.58(2) & 0.5 \\
              & $24$ & 1.54(4) & 0.4 \\
              & $32$ & 1.57(6) & 0.4 \\
\hline
                          & $8$ & 1.51(3) & 1.4 \\
$\chi_{\rm loc}(x_\perp=2)$ & $12$ & 1.37(6) & 1.1 \\
                          & $16$ & 1.36(11) & 1.1  \\
                          & $24$ & 1.3(3) & 1.2 \\
\hline
                          & $8$ & 1.71(2) & 2.8 \\
$\chi_{\rm loc}(x_\perp=2)$ & $12$ & 1.52(4) & 1.3 \\
                          & $16$ & 1.37(7) & 1.0  \\
                          & $24$ & 1.0(2) & 0.7 \\
\end{tabular}
\end{ruledtabular}
\label{table.fitxO}
\end{table}
In order to investigate the SDE of the magnetization at the normal fixed point, we have also sampled the order-parameter profile for $h_0=\infty$, $K=K_c$, and lattices sizes $L$ as above. We have analyzed the magnetization at a fixed distance $x_\perp\ll L$, as a function of $L$. According to Eq.~(\ref{profcrit_SDE_fit}), in this regime the dependence on $L$ of $m(x_\perp)$ allows to extract the scaling dimension $x_O$ of the leading operator in the SDE of Eq.~(\ref{SDE_approx}). We have fitted $m(x_\perp)$ to
\begin{equation}
m(x_\perp) = A L^{-x_O} + B,
\label{fitxO_m}
\end{equation}
leaving $A$, $x_O$, and $B$ as free parameters, and fixing $x_\perp=2$, $3$, close to the pinning-field line. In Table \ref{table.fitxO}, we report the fit results as a function of the minimum lattice size $L_{\rm min}$ used.

In order to critically assess the reliability of the fit results, it is important to recall that the scaling behavior discussed in Sec.~\ref{sec:fss:op} is valid for $x_\perp \gtrsim \sigma_0$, with $\sigma_0=O(1)$ a nonuniversal length that governs the short-distance nonuniversal behavior. Therefore a determination of $x_O$ via fits of Eq.~(\ref{fitxO_m}) must be repeated for increasing values of $x_\perp$, in order to monitor possible residual, short-distance, nonuniversal effects. With these regards, we mention that fit results for $x_\perp=1$ (not shown here), display a significantly large $\chi^2/{\rm DOF}\approx 1.4-1.7$ for $L_{\rm min}=8\ldots 32$, and a fitted value of $x_O\approx 1.4-1.5$, which exhibits a small but statistically significant discrepancy with the results of Table \ref{table.fitxO}. Moreover, the argument that leads to Eq.~(\ref{profcrit_SDE_fit}) is valid only for $x_\perp/L\ll 1$. Therefore, for a given value of $x_\perp$, in order to assess the reliability of  fitted values of $x_O$, it is necessary to monitor the stability of the results for increasing values of $L_{\rm min}$. The condition $x_\perp/L\ll 1$ can also be seen as a limitation in a numerical determination of $x_O$, because for increasing values of $x_\perp$ larger lattices sizes are required, in order to satisfy $x_\perp/L\ll 1$; on the same time, an increasing precision in the MC data is needed to fit Eq.~(\ref{fitxO_m}), because $m(x_\perp)$ decreases in magnitude for increasing values of $x_\perp$ [eventually, for $L\rightarrow \infty$, $m(x_\perp)\propto x_\perp^{-\beta/\nu}(1 + x_\perp^{-\omega}g_{pc,\infty})$, see Eq.~(\ref{profcrit_infL})]. With limited available lattice sizes, this allows to consider only a few values of $x_\perp$.

Along with these considerations, we have critically inspected the fit results of Table \ref{table.fitxO}. For $x_\perp=2$, a good $\chi^2/{\rm DOF}$ is found for $L_{\rm min}\ge 12$. However, the fitted value of $x_O$ for $L_{\rm min}=12$ is only in marginal agreement with the corresponding result for $L_{\rm min}=16$, suggesting that data at $L=12$ may be still affected by subleading corrections to Eq.~(\ref{fitxO_m}). For $L_{\rm min}\ge 16$, the fitted values of $x_O$ are stable and in mutual agreement, suggesting that the condition $x_\perp/L\ll 1$ is effectively satisfied for $x_\perp/L \lesssim 2/16=1/8$. A similar pattern in the fit results is observed for $x_\perp=3$, where a good $\chi^2/{\rm DOF}$ is found for $L_{\rm min}\ge 12$, but the fitted value of $x_O$ for $L_{\rm min}=12$ shows a small, significant deviation with respect to to the results for $L_{\rm min}\ge 16$. For $x_\perp=3$, the aforementioned condition $x_\perp/L\lesssim 1/8$ is satisfied for $L_{\rm min}\ge 24$, whose results are also in perfect agreement with the values found for $x_\perp=2$. By judging conservatively the results of Table \ref{table.fitxO} we infer
\begin{equation}
x_O=1.52(6), \qquad \eta_\parallel=2.04(12),
\label{xO}
\end{equation}
where $\eta_\parallel$ is related to $x_O$ by Eq.~(\ref{etapar}).
As shown in Eq.~(\ref{corr_par}), the scaling dimension $x_O$ enters also in the asymptotic behavior of the two-point function along the pinning-field line, which decays with an exponent $2x_O=3.0(1)$. Therefore the critical exponents $x_O$ and $\eta_\parallel$ characterize the critical adsorption UC on a line.

The existence of a line operator with a scaling dimension given in Eq.~(\ref{xO}) implies the presence of a scaling field with RG dimension $y_O=1-x_O=-0.52(6)$. Such an irrelevant scaling field enters in the FSS ansatz of the singular part of the free energy density, Eq.~(\ref{fsingBC}), as well as in any critical quantity, such as the magnetization $m$, giving rise to scaling corrections $\propto L^{-0.52(6)}$. This value of $\omega$ matches very well the value of $\omega$ given in Eq.~(\ref{omega}), which is obtained by fitting the scaling corrections of $m$. It indicates that the unexpectedly slowly decaying scaling corrections found in this model originate from the line operator $O$. Indeed, for the improved Blume-Capel model of Eq.~(\ref{HBC}), the leading irrelevant bulk scaling field is suppressed, while analytical scaling corrections originating from the broken translational invariance have $\omega=1$. Therefore the slowly decaying scaling corrections found here must be attributed to the presence of a line irrelevant scaling field.

To further strengthen this picture, we have also analyzed the FSS behavior of the local susceptibility $\chi_{\rm loc}(x_\perp)$ at $h_0=\infty$ and $K=K_c$. As for $m(x_\perp)$, we fix $x_\perp$ close to the pinning-field line, and study the $L$-dependence. According to Eq.~(\ref{chiloc_fss}), in this regime the leading FSS behavior is equivalent to the one for the local magnetization. Thus we have fitted $\chi_{\rm loc}(x_\perp)$ to
\begin{equation}
\chi_{\rm loc}(x_\perp) = A L^{-x_O} + B,
\label{fitxO_chiloc}
\end{equation}
leaving $A$, $x_O$, and $B$ as free parameters. Fit results are reported in Table \ref{table.fitxO}. Concerning the reliability of the fitted results, the same considerations outlined for $m(x_\perp)$ hold in this case. We observe that the fit results displays a lower quality as compared to the ones for the local magnetization. A good $\chi^2/{\rm DOF}$ is obtained only in a few cases, and the fitted values of $x_O$ exhibits a dependence on $L_{\rm min}$ which is larger than the one found for $m(x_\perp)$, suggesting that the ansatz of Eq.~(\ref{fitxO_chiloc}) does not fully describes the data. Such a difficulty can be traced back to the existence of a correction to the leading FSS behavior. In fact, as discussed in Sec.~\ref{sec:fss:op}, on the top of the leading FSS behavior of $\chi_{\rm loc}(x_\perp)$ given by Eq.~(\ref{chiloc_fss}), there is also a contribution $\propto L^{-\eta_\parallel}$. Such a term constitutes a subleading $L$-dependence, which is suppressed with respect to to the leading $L$-dependence $\propto L^{-x_O}$ by a factor $\propto L^{-\eta_\parallel+x_O}=L^{-(x_O-1)}$. Therefore the neglected contribution $\propto L^{-\eta_\parallel}$ acts as a correction to scaling, with an effective exponent $\omega=x_O-1=0.52(6)$, where we have used the estimate of Eq.~(\ref{xO}). As for the case of the full magnetization $m$, this constitutes a slowly decaying correction to scaling, which is responsible for the lower quality of fits found. Our data are not precise enough to fit a FSS ansatz of the form $A L^{-x_O} + B + CL^{-\eta_\parallel}$, which includes the correction. Despite these limitations, we observe that the fitted values of $x_O$ extracted from $\chi_{\rm loc}$ display only a small deviation with respect to the estimate of Eq.~(\ref{xO}) and substantially confirm the result obtained from $m(x_\perp)$. Moreover, the fitted values of $x_O$ obtained from $\chi_{\rm loc}$ confirm the analysis of Sec.~\ref{sec:fss:op} and Eq.~(\ref{chiloc_fss}), according to which the leading FSS behavior of $\chi_{\rm loc}(x_\perp)$ is characterized by the exponent $x_O$ and not by $\eta_\parallel$, as usual.

Reference [\onlinecite{Allais-14}] studied the three-dimensional Ising model in the presence of an external field coupled to a defect line. In Ref.~[\onlinecite{Allais-14}], the magnetization profile is first extrapolated to the TDL, and then fitted to $c_1/x_\perp^\Delta+c_2/x_\perp^{\Delta_2}$, with $\Delta=0.526(5)$ and $\Delta_2=0.93(3)$. According to Eq.~(\ref{profcrit_infL}) and the discussion of Sec.~\ref{sec:fss:op}, the leading decay exponent $\Delta$ corresponds to the magnetization exponent $\beta/\nu$, whereas $\Delta_2-\Delta$ is identified with the leading correction-to-scaling exponent $\omega$. Within the precision quoted by Ref.~[\onlinecite{Allais-14}], the fitted value of $\Delta$ is in marginal agreement with $\beta/\nu=0.51813(5)$~\cite{Hasenbusch-10}, while the difference $\Delta_2-\Delta=0.40(4)$ shows a discrepancy of about two combined error bars with our determined value of $\omega$ [Eq.~(\ref{omega})].

\subsection{Bilayer Heisenberg model}
\label{sec:results:heisenberg}
\begin{figure}[t]
\centering
\includegraphics[width=\columnwidth]{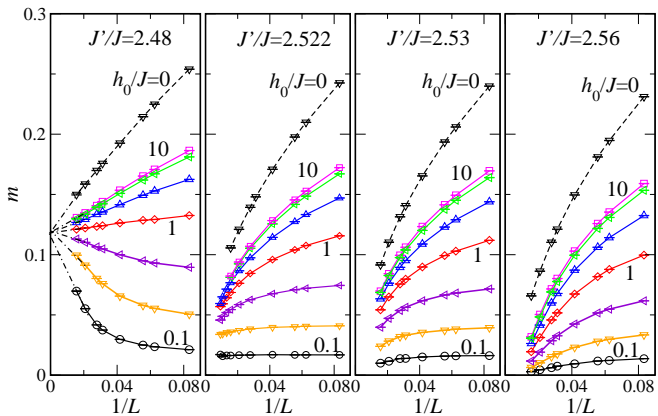}
\caption{Linear system size ($L$) dependence of the order-parameter estimate $m$ for pinning fields of different magnitude $h_0/J=0.1,0.25,0.5,1,2,5,10$ (bottom to top) as well as obtained at $h_0=0$ from the structure factor (top curve) for four different values of the coupling ratio $g$ near the quantum critical point, as indicated. Dashed-dotted lines in the left-most panel indicate linear extrapolations to the TDL.}
\label{fig:figbhmvs1oL}
\end{figure}

In the following, we present our results from QMC simulations both for $h_0=0$, using the conventional order-parameter estimate in Eq.~(\ref{eq:mconv}), as well as for finite values of $h_0$, based on the estimator in Eq.~(\ref{eq:mnew}). 
We employ stochastic series expansion QMC simulations~\cite{Sandvik99,Syljuasen02,Alet05} for square lattice bilayers of a linear extent $L$ ranging between  12 and  96 and periodic boundary conditions, containing $2L^2$ spins.  The temperature was scaled with the linear system size as $1/T=2 L$ in order to target the quantum critical regime. From previous simulations~\cite{Wang06}, the critical coupling ratio has been estimated as $g_c=2.5220(1)$. A first issue that we consider is, how the presence of a pinning field $h_0$ affects the behavior of the order-parameter estimate near  the quantum critical point, and how the position  of the quantum critical point ($g_c$) may be estimated based on pinning-field data. 

In Fig.~\ref{fig:figbhmvs1oL}, the order-parameter estimate $m$ is shown as a function of the inverse linear system size $1/L$ for different values of the coupling ratio $g=J'/J$ and magnitudes $h_0$ of the pinning field, using the estimator in Eq.~(\ref{eq:mnew}) for finite values of $h_0$ and  the  conventional estimator in  Eq.~(\ref{eq:mconv})  in the absence of the pinning field, $h_0=0$. We observe that for $g=2.48$, which resides close to criticality but still inside the ordered phase, the data for $m$ from the larger systems allows for a linear extrapolation in $1/L$ to an essentially $h_0$-independent value of the order-parameter estimate in the TDL. Moreover, the data  at $g=2.48$ for small values of $h_0$ exhibit a characteristic increase of $m$ with increasing  system size. For $g=2.53$ and $g=2.56$, all data  for different values of $h_0$ exhibit a reduction of $m$ upon increasing the system size; however, an extrapolation to the TDL  would need to account for the non-linear-behavior in the $1/L$-dependence of $m$ in these cases. The data for $g=2.522$, which essentially resides at the quantum critical point, show similar difficulties for an extrapolation to the value in the TDL, with the data for $h_0/J=0.1$ being $L$-independent within the accessed range of system sizes. 

\begin{figure}[t]
\centering
\includegraphics[width=\columnwidth]{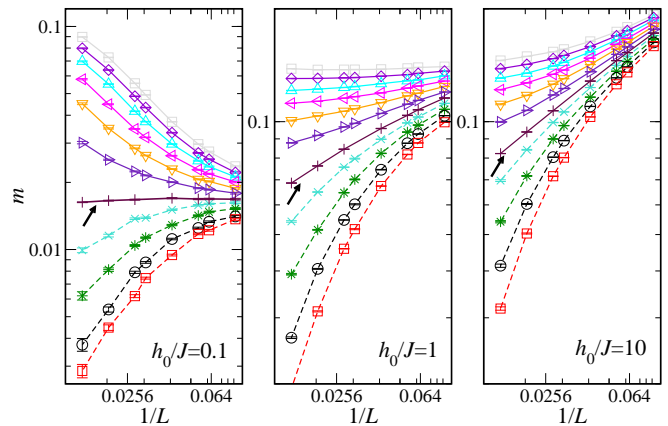}
\caption{Linear system size ($L$) dependence of the order-parameter estimate $m$ for pinning fields of different magnitude $h_0/J=0.1, 1,10$ (left to right) in a double logarithmic plot for various values of the coupling ratio $g=2.46, 2.47, 2.48, 2.49, 2.5, 2.51, 2.522, 2.53, 2.54, 2.55, 2.56$ (top to bottom) near the quantum critical point, with the data for $g=2.522$ indicated by arrows. Data within the ordered (disordered) region are connected by solid (dashed) lines.}
\label{fig:figbhlogmvs1oL}
\end{figure}

An estimate of the quantum critical point can nevertheless be obtained from the pinning-field data via the identification of a leading algebraic scaling of the data at the critical coupling ratio: the finite-size data taken away from  the critical point eventually exhibits a bending for large system sizes; if the coupling $g$ is fixed to  its quantum critical value $g_c$ though, we instead observe an algebraic decay for the  larger system sizes. This qualitative behavior is insensitive to the value of $h_0$, as shown in Fig.~\ref{fig:figbhlogmvs1oL} for different values of $h_0$ in a double-logarithmic plot. Here, we also observe that the asymptotic slope in the double-logarithmic plot at the quantum critical coupling  depends on the value of $h_0$. We  relate this to the crossover behavior discussed in Sec.~\ref{sec:fss}, from the $h_0=0$ fixed point to the infinite $h_0$ fixed point that we analyze in more detail below. 

\begin{figure}[t]
\centering
\includegraphics[width=\columnwidth]{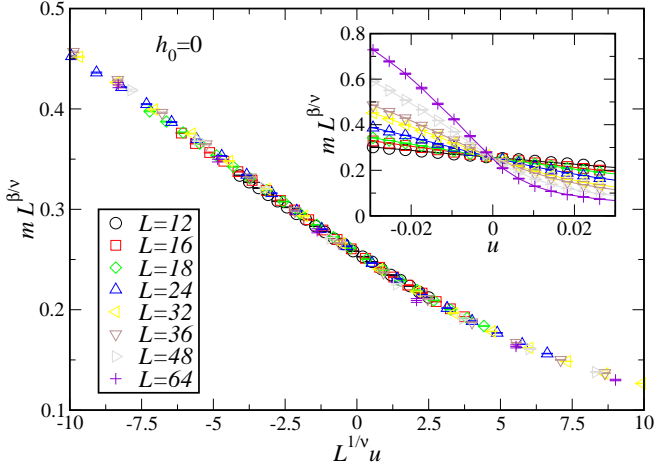}
\caption{Data collapse and crossing (inset) plot of the order-parameter estimate for the bilayer spin-1/2 model in the absence of pinning field, $h_0=0$.}
\vspace{2em}
\label{fig:figbh0collapse}
\end{figure}

Before we turn to a more quantitative analysis of the finite-size scaling behavior in the presence of a pinning field, 
we want to demonstrate explicitly  that  the data for $m$ obtained for finite values of the pinning field $h_0$ does not allow to extract the critical exponents of the quantum critical point based on the conventional leading finite-size analysis that is feasible in the absence of the pinning field, $h_0=0$, using, e.g., the structure factor-based estimate for $m$.
In particular, the data obtained in the conventional ($h_0=0$) approach exhibit a robust crossing point at $u=0$ when plotted as $mL^{\beta/\nu}$ versus $u=(g-g_c)/g_c$, c.f. the inset of Fig.~\ref{fig:figbh0collapse} . Furthermore, we obtain a good data collapse upon plotting $m L^{\beta/\nu}$ versus $uL^{1/\nu}$, as anticipated from the leading finite size scaling behavior,  shown in the main panel of Fig.~\ref{fig:figbh0collapse}. Here, we employed the values of the critical exponents $\beta/\nu=0.5188(3)$ and $\nu=0.7112(5)$ for the three-dimensional Heisenberg UC~\cite{CHPRV-02}, while it is also feasible to obtain appropriate estimates for these exponents, e.g., based on our data for $m$ for $L\ge 36$, we obtain the estimates $\beta/\nu=0.515(3)$ and $\nu=0.70(1)$ from an unbiased fit. For a more extended analysis of the QMC data in the absence of the pinning field, we refer to Ref.~[\onlinecite{Wang06}] and now focus on the case of a finite pinning field. 

\begin{figure}[t]
\centering
\includegraphics[width=\columnwidth]{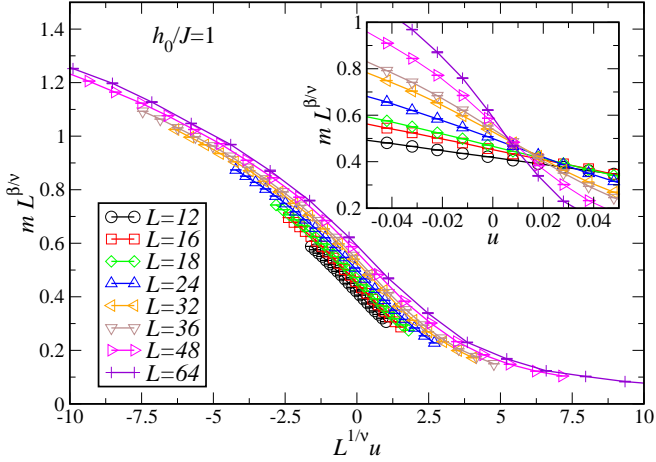}
\caption{Attempted data collapse (main panel) and crossing (inset) plot of the order-parameter estimate for the bilayer spin-1/2 model with a  pinning field of strength $h_0=J$.}
\label{fig:figbh1collapse}
\end{figure}

The data obtained for finite values of $h_0$ clearly contrast to the behavior observed in Fig.~\ref{fig:figbh0collapse}. This is illustrated in
Fig.~\ref{fig:figbh1collapse}  for $h_0=J$, employing the reference values of the critical exponents~\cite{CHPRV-02} .
In this case, the crossing points in the insets exhibit pronounced drifts and the data for different system sizes fails to  collapse in the main panel. We note that we also attempted to perform unbiased fits of the data at a  given value of $h_0\neq 0$ to the above scaling ansatz with $\beta$ and $\nu$ as free parameters. We observed that (i) one may still collapse the data based on visual inspection, but (ii) the resulting values of $\chi^2/{\rm DOF}$ are actually rather large (ranging, depending on the value of $h_0$, between 10 and several hundreds)  thus indicating that these collapses are in fact not good, and (iii) the thereby obtained estimates for the critical exponents vary strongly with the value of $h_0$ as well as the considered range of system sizes. The estimates for the critical exponents are furthermore found to approach closer to the true values for larger values of $h_0$, but even for a very large value of $h_0/J=1000$, which effectively corresponds to the infinite $h_0$ limit, we observe systematic deviations.  These observations are in accord with our general scaling analysis in Sec.~\ref{sec:fss}: the presence of the pinning field leads to a crossover behavior from the $h_0=0$ fixed point to the infinite-$h_0$ fixed point, as well as to important scaling corrections that in effect require an extended scaling analysis even to only approximately obtain  reliable estimates for the critical exponents. 

\begin{figure}
\centering
\includegraphics[width=\columnwidth]{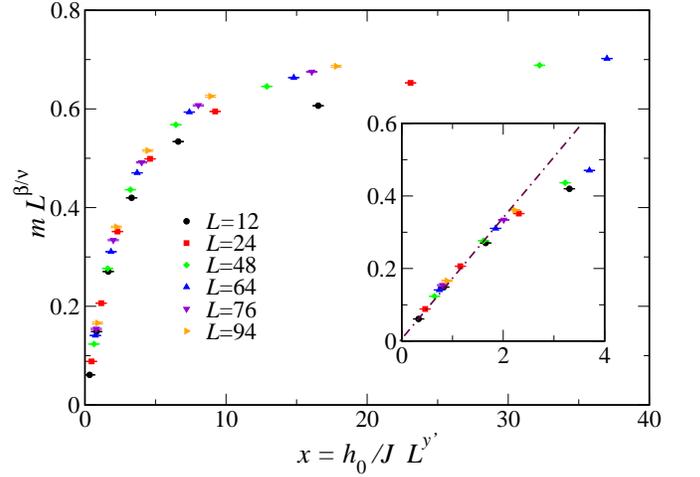}
\caption{Scaling collapse of the order-parameter estimate $m$ for the Heisenberg bilayer model at the quantum critical coupling $g_c=2.522$~\cite{Wang06}, and for finite values of the pinning field $h_0$. The critical exponents are those of the 3D Heisenberg UC, with $y'=(1-\eta)/2=0.4813(3)$ [see Eq.~(\ref{yprime})] and $\beta/\nu=0.5188(3)$~\cite{CHPRV-02}. Up to scaling corrections,  which are clearly visible for the smaller system sizes, $mL^{\beta/\nu}$ scales as a function of $x=h_0/J L^{y'}$.
The inset shows the low-$x$ data on a linear scale, with the dash-dotted line indicating the linear scaling at low values of $x$. 
}
\label{fig:figbhscaling}
\end{figure}

To further analyze the impact of these scaling corrections in light of our general discussion in Sec.~\ref{sec:fss}, we next analyze the FSS of the data for $m$ for different values of $h_0$ as obtained from simulations performed at the estimate $g=g_c=2.522$ for the quantum critical point. 
From the general FSS analysis in Sec.~\ref{sec:fss}, we expect that up to scaling corrections, 
 $mL^{\beta/\nu}$ is a function of the scaling variable $h_0L^{y'}$, c.f. Eq.~(\ref{mcrit}). 
Based on the 
value of the critical exponent $\eta=0.0375(5)$  of the three-dimensional Heisenberg UC~\cite{CHPRV-02}, we obtain $y'=0.4813(3)$ and $\beta/\nu=0.5188(3)$. In Fig.~\ref{fig:figbhscaling} we plot $mL^{\beta/\nu}$ as a function of $h_0L^{y'}$, obtaining a good collapse of the QMC data, which supports Eq.~(\ref{mcrit}) with $y'$ given in Eq.~(\ref{yprime}). Scaling corrections however are  visible  for the smaller system sizes. 
The data is also in accord with  Eq.~(\ref{mcrit_norm_asymptotic}), as  $mL^{\beta/\nu}$ tends to approach a constant value for large values of the scaling variable $\widetilde{h_0}=h_0L^{y'}$. In line with the discussion of Sec.~\ref{sec:fss:h0} and our findings for the classical model, Fig.~\ref{fig:figbhscaling} again illustrates a crossover between the $h_0=0$ and the $h_0=\infty$ fixed points.
For small values of $\widetilde{h_0}=h_0L^{y'}$, the magnetization scaling function at fixed aspect ratio $\rho$ $f_{mc}(\widetilde{h_0}) \equiv f_{mc}(\widetilde{h_0},\rho=0.5/c_0)$ on the right-hand side of Eq.~(\ref{mcrit}) can be approximated by its first-order Taylor expansion, such that, neglecting scaling corrections, $m \simeq L^{-\beta/\nu}f_{mc}'(0)h_0L^{y'}=f_{mc}'(0)h_0L^{-\eta}$. This means that for a small pinning field $h_0$ there is a range in lattice sizes where the magnetization approximately scales as $\propto L^{-\eta}$; such a range is defined by the interval in $h_0L^{y'}$ where the scaling function shown in Figs.~\ref{fig:scalingmBC}, \ref{fig:figbhscaling} can be approximated by a straight line. Since $\eta=0.0375(5)$~\cite{CHPRV-02}, this accounts for the observed weak $L$-dependence of the magnetization at $g=g_c$ and $h_0=0.1$, as shown in the second panel of Fig.~\ref{fig:figbhmvs1oL} and in the first panel of Fig.~\ref{fig:figbhlogmvs1oL}. We emphasize that this is only a transient behavior, since $m\propto L^{-\beta/\nu}$ for $L\rightarrow\infty$, as illustrated by the flat tail of the scaling functions shown in Figs.~\ref{fig:scalingmBC} and Fig.~\ref{fig:figbhscaling} [see Eq.~(\ref{mcrit_norm_asymptotic})].

We next analyze the scaling behavior of $m$ at the  $h_0=\infty$ fixed point and estimate the relevance of the leading scaling corrections. For this purpose, we employ the data taken at $g=g_c=2.522$ for $h_0/J=1000$, which effectively corresponds to the $h_0=\infty$ limit.

First, we consider fits of the data for $m$ to  the leading FSS form, Eq.~(\ref{fitm}), leaving $A$ and $\beta/\nu$ as free parameters.
In Table \ref{table.bhfitm} we report the fit results as a function of the minimum lattice size $L_{\rm min}$ taken into account. The fitted value of $\beta/\nu$ show a significant deviation from the expected value $\beta/\nu=0.5188(3)$~\cite{CHPRV-02}, with large values of $\chi^2/{\rm DOF}$.  Furthermore, the results exhibit a systematic drift which is compatible with a slow approach towards the expected value.

\begin{table}
\caption{Fits of the magnetization $m$ at $h_0=1000$ to Eq.~(\ref{fitm}), for $L=12\ldots 64$. $L_{\rm min}$ indicates the minimum lattice size $L$ taken into account. For the three-dimensional Heisenberg UC $\beta/\nu=0.5188(3)$~\cite{CHPRV-02}.}
\begin{ruledtabular}
\begin{tabular}{l..}
\multicolumn{1}{c}{$L_{\rm min}$} & \multicolumn{1}{c}{$\beta/\nu$} & \multicolumn{1}{c}{$\chi^2/{\rm DOF}$} \\
\hline
$12$ & 0.4406(1) & 506 \\
$16$ & 0.4461(2) & 161 \\
$18$ &  0.4484(2) & 125\\
$24$ & 0.4538(3) & 36 \\
$32$ & 0.4596(5) & 3.2 \\
\end{tabular}
\end{ruledtabular}
\label{table.bhfitm}
\end{table}

The results of Table \ref{table.bhfitm}  hint to the presence of important scaling corrections at the $h_0=\infty$ fixed point, similar to what was observed in the classical model. 
In the presence of corrections to scaling with a leading exponent $\omega$, Eq.~(\ref{fitm}) is modified to Eq.~(\ref{fitm_omega}), as discussed there.  
We thus fitted the data of $m$ against the right-hand side of Eq.~(\ref{fitm}), leaving $A$, $B$, $\beta/\nu$, and $\omega$ as free parameters. In Table \ref{table.bhfitm_omega}, we report the results of these fits. For every minimum lattice size $L_{\rm min}$ considered, $\chi^2/{\rm DOF}$ is small and the results give a value of $\beta/\nu$ in reasonable agreement with the value of the three-dimensional Heisenberg UC $\beta/\nu=0.5188(3)$~\cite{CHPRV-02}. The fits provide a significantly small value of $\omega\approx 0.4-0.6$, in line with the observation of the presence of slowly decaying scaling corrections also in the present case. In order to obtain a more precise estimate of $\omega$, we have repeated the fits of $m$ to Eq.~(\ref{fitm_omega}), fixing $\beta/\nu$ to the value of the Heisenberg UC $\beta/\nu=0.5188(3)$~\cite{Hasenbusch-10}. The corresponding results are also reported in Table \ref{table.bhfitm_omega}. The fitted value of $\omega$ is now more stable, and the $\chi^2/{\rm DOF}$ similarly small. A conservative estimate in this case is $\omega=0.32(4)$. This correction exponent is  smaller than the value we estimated for the three-dimensional Ising UC. In addition to the different UC, we would like to remark also, that for the classical model, we were able to access system sizes that are an order of magnitude larger than in the present case. The important conclusion to be drawn from the present analysis however is that the presence of the pinning field leads to only-weakly decaying scaling corrections, which need to be accounted for in the estimation of critical exponents from pinning-field data of the order-parameter estimate.
Like in the Blume-Capel model, such additional scaling corrections must be attributed to an irrelevant line scaling field, because the leading corrections to scaling in the absence of a pinning fields decays with a significantly larger exponent $\omega=1.0(3)$~\cite{Wang06}.

\begin{table}
\caption{Fit of the magnetization $m$ at the critical adsorption fixed point $h_0=1000$ to Eq.~(\ref{fitm_omega}) for $L=12\ldots 64$. $L_{\rm min}$ indicates the minimum lattice size $L$ taken into account. In the first set of fits (above), $\beta/\nu$ is a free parameter. In the second set (below), $\beta/\nu$ is fixed to the Heisenberg UC value $\beta/\nu=0.5188$~\cite{CHPRV-02}.}
\begin{ruledtabular}
\begin{tabular}{l@{}.@{}.@{}.}
\multicolumn{1}{c}{$L_{\rm min}$} & \multicolumn{1}{c}{$\beta/\nu$} & \multicolumn{1}{c}{$\omega$} & \multicolumn{1}{c}{$\chi^2/{\rm DOF}$} \\
\hline
$12$ & 0.49(1) & 0.5(1) & 0.31 \\
$16$ & 0.51(3) & 0.4(3) & 0.35  \\
$18$ & 0.50(3) & 0.6(4) & 0.38 \\
\hline
$12$ & & 0.345(8) & 0.67 \\
$16$ & & 0.32(1) & 0.28 \\
$18$ & & 0.32(2) & 0.36\\
$24$ & & 0.32(4) & 0.47 \\
\end{tabular}
\end{ruledtabular}
\label{table.bhfitm_omega}
\end{table}

\section{Discussion and future directions}\label{sec:conclusions}
In this paper we have analyzed the critical behavior in the presence of a pinning field. We have studied the two-dimensional bilayer Heisenberg model in the presence of a pinning field coupled to a single site, and the three-dimensional improved Blume-Capel model with a pinning field coupled to a line. Although these models realize different O(N) UCs, their critical behavior is qualitatively similar because, under the quantum-to-classical mapping, the bilayer Heisenberg model becomes equivalent to a three-dimensional model, in the presence of an external field coupled to a line.

Compared to QMC simulations of the bilayer Heisenberg model, classical MC simulations of Blume-Capel model allow to obtain significantly larger lattice sizes and improved statistical precision, thus permitting a deeper analysis of the finite-size scaling properties of the model. Another crucial difference between the models lies in the scaling corrections: the Blume-Capel model considered here is an improved model, where the amplitude of the leading irrelevant scaling field is suppressed. Improved models have turned out to be instrumental in high-precision investigations of critical phenomena~\cite{PV-02,Hasenbusch-10,Hasenbusch-10c,PTD-10,Hasenbusch-11,PTTD-13,PTTD-14,Hasenbusch-14,PT-13,Hasenbusch-08,Hasenbusch-12,CHPRV-02}, since controlling scaling corrections is essential for a reliable computation of universal critical properties, such as critical exponents. 

The RG flow of the models exhibits important analogies with the critical behavior at surfaces~\cite{Binder-83,Diehl-86,Diehl-97}. In particular, one distinguishes between ``bulk'' and ``line'' couplings. While the RG flow of the bulk couplings is independent of the line ones, the RG flow of the latter depends also on the bulk couplings. As a consequence, like in surface critical phenomena, a given bulk UC splits into different line UCs.
According to the RG analysis of Sec.~\ref{sec:fss}, at the bulk fixed point the RG flow of the line couplings has a fixed point for zero pinning field $h_0$, which is the analog of the ordinary UC in surface critical phenomena. At this fixed point, the pinning field is a relevant perturbation, whose scaling dimension can be exactly determined as a function of the bulk critical exponents. In line with field-theory calculations~\cite{Hanke-00} and analogous to surface critical phenomena, for $h_0\ne 0$ the models exhibit a crossover to a critical adsorption, or normal, fixed point $h_0=\infty$. By means of MC simulations, we have checked this scenario for both the bilayer Heisenberg model and the improved Blume-Capel model, and verified the exact result for the scaling dimension of the pinning field at the $h_0=0$ fixed point.
This picture is also expected to hold for the Hubbard model on the honeycomb lattice, which undergoes a quantum phase transition in the Gross-Neveu Heisenberg UC~\cite{HJV-09,PTHAH-14,OYS-16}. For this model the scaling dimension of the pinning field is found to be significantly smaller than for the O(N) models, such that very large lattice sizes would be needed in order to reach the asymptotic behavior; see the discussion in Appendix \ref{sec:honeycomb}.

At the critical adsorption fixed point $h_0=\infty$, we observe unexpected slowly decaying scaling corrections, which originate from an irrelevant line scaling field. Indeed, as we discuss in Sec.~\ref{sec:fss:op}, the decay of the order parameter close to the pinning-field line is characterized by a finite-size correction, which is the analog of the so-called distant-wall corrections in surface critical phenomena~\cite{FG-78}. However, unlike the latter case, and in line with field-theory results~\cite{Hanke-00}, such a finite-size correction is characterized by a new exponent, originating from a presently unknown line operator. This exponent enters also in the leading decay of the two-point function along the pinning-field line, thereby characterizing the critical adsorption UC. The MC results of the improved Blume-Capel model support the identification of the corresponding line scaling field with the leading irrelevant perturbation, which accounts for the observed slowly decaying scaling corrections. We expect such a picture to hold also for the bilayer Heisenberg model. Indeed, the observed emergence of unusual slowly decaying scaling corrections must originate from a line irrelevant scaling field, whose corresponding line operator is expected to enter in the short-distance expansion of the order parameter close to the pinning-field line. A candidate for this unknown line operator is the $\phi^3$ operator which, unlike its bulk counterpart, is not redundant~\cite{CD-90,DC-91}; its RG flow has been studied by means of field theory in Refs.~[\onlinecite{CD-90,DC-91}] for a semi-infinite three-dimensional geometry bounded by a two-dimensional surface, and in the presence of a surface field. Incidentally, the scaling dimension of the operator $\phi^3$ at the Gaussian fixed point is $3/2$, close to the result of Eq.~(\ref{xO}).

In view of these results, a correct FSS analysis of models in the presence of a finite pinning field must include a corresponding scaling field, which is responsible for the crossover behavior between the $h_0=0$ and the $h_0=\infty$ fixed points. Even at the critical adsorption fixed point $h_0=\infty$, particular care has to be taken in the analysis of the critical exponents, since the model is affected by slowly decaying scaling corrections. On the other hand, away from the critical point, as we show in Sec.~\ref{sec:results:heisenberg}, the pinning-field approach allows to identify the phases of the model. Nevertheless, in the vicinity of the bulk critical point, the extrapolation to the thermodynamic limit of the order parameter presents some difficulties, due to the aforementioned crossover behavior.

In the case of the three-dimensional Blume-Capel model, the system studied here can be experimentally realized by considering a classical binary liquid mixture. At the critical demixing point, the mixture undergoes a second-order phase transition in the Ising UC, where the order parameter is given by the difference of the concentration of one of the species with its critical concentration. The inclusion of a colloidal particle in the solvent typically leads to a preferential adsorption of one component of the mixture at the colloidal surface, which can be modeled by a surface field~\cite{FG-78}. Thus, the present setup of a pinning-field line can be experimentally realized by considering an elongated cylindrical colloid immersed in a critical binary mixture, in the limit of small radius and large length of the colloid. A variety of rodlike particles are experimentally available, such as Boehmite rods~\cite{BL-93,BPL-94}, gold nanorods~\cite{PJPSLMM-05}, carbon nanotubes~\cite{Iijima-91}, microtubules~\cite{GMNH-93}, lipid tubules~\cite{ZF-06}, the mosaic tobacco virus~\cite{NS-86}, and cylindrical micelles~\cite{GGWCMRWM-10}.

The present setup lends itself to further generalizations.

(i) Surface pinning field.

Here we consider a pinning field coupled to a one-dimensional line in a $d$-dimensional quantum model or, correspondingly, to a two-dimensional plane in a classical model in $D=d+1$ dimensions.

This is a case that closely resembles surface critical phenomena. A simple generalization of the argument of Sec.~\ref{sec:fss:h0} allows to conclude that the RG dimension of the pinning field at the $h_0=0$ fixed point is $y'=y_h-(d-1)$ for the quantum model, and $y'=y_h-(D-2)$ for the classical one. For the 3D O(N) UC, one has $y'\approx 1.5$, therefore, as for the models studied in this paper and in surface critical phenomena, such pinning field is a relevant perturbation. Analogous to the present case, the RG flow for $h_0>0$ leads to a critical adsorption fixed point at $h_0=\infty$. We observe that for the 3D $O(N)$ UC the value of the exponent $y'\approx 1.5$ is considerably larger than the one found for a pinning-field line investigated in this paper. Therefore, there is a significantly faster crossover from the ``ordinary'' fixed point $h_0=0$ to the normal UC $h_0=\infty$. In other words, the flattening of the scaling functions shown in Figs.~\ref{fig:scalingmBC} and \ref{fig:figbhscaling} occurs for smaller lattice sizes and pinning-field strengths.

At the critical adsorption fixed point, one expects the same critical behavior as for the case of an ordered surface, i.e., a surface normal UC. In this case, the leading operator in the short-distance expansion of the order parameter is the perpendicular component of the stress-energy tensor~\cite{Cardy-90,ES-94}, such that the leading finite-size correction in the decay of the order parameter is $\propto L^{-3}$, and the correlations parallel to the plane decay as $\propto |\vec{x}_\parallel'-\vec{x}_\parallel|^{-2D}$ (see the discussion at the end of Sec.~\ref{sec:fss:op}). Concerning irrelevant surface operators, field theory calculations for the Ising UC in the presence of a surface field reported an additional irrelevant surface scaling field with negative dimension $\omega=\varepsilon$, thereby suggesting the existence of additional scaling corrections of relevant magnitude~\cite{CD-90,DC-91}. However, MC studies of improved models in the Ising UC did not find additional scaling corrections~\cite{PTD-10,Hasenbusch-10c,Hasenbusch-11,Hasenbusch-12,PTTD-13,PTTD-14} beside those $\propto L^{-1}$ arising from the broken translational invariance. Therefore, all critical observables are affected by corrections to scaling $\propto L^{-\omega}$, with $\omega$ the leading bulk irrelevant exponent ($\omega \approx 0.8$~\cite{CHPRV-02,Hasenbusch-10} for the 3D O(N) UC), or $\propto L^{-1}$ in the case of an improved model, where the leading bulk irrelevant scaling field is suppressed. Nevertheless, the magnetization is further affected by a significantly larger correction arising from the nonsingular part of the free energy. At the bulk critical point and the surface critical adsorption fixed point, neglecting irrelevant operators and the corrections $\propto L^{-1}$, the magnetization $m$ satisfies the following FSS ansatz
\begin{equation}
m=AL^{-\beta/\nu} + \frac{1}{L^{D-2}}B,
\label{mcrit_surfacepf}
\end{equation}
where $B$ is the amplitude of the nonsingular part of the surface magnetization [compare with Eq.~(\ref{mcrit})]. Equation (\ref{mcrit_surfacepf}) shows that the nonsingular part of the free energy gives rise to a correction to scaling with an effective exponent $\omega=D-2-\beta/\nu$. For the three-dimensional O(N) UC one has  $\omega\approx 0.5$, hence this constitutes a significant scaling correction. In absence of additional irrelevant surface fields, this is the leading correction to scaling. An advantage with respect to the case of a pinning-field line is that the leading $\omega$ exponent is not a new exponent, but is obtained from the magnetization exponent $\beta/\nu$. Notice that the background scaling corrections in Eq.~(\ref{mcrit_surfacepf}) affect also the FSS behavior of the magnetization in the presence of a finite pinning field.

Away from the critical point, a surface pinning field can be used to identify the ordered and disordered phases, as done in Sec.~\ref{sec:results:heisenberg}. Nevertheless, one expects in the paramagnetic phase a slower convergence of the magnetization to the thermodynamic limit of $0$, because of the relatively larger contribution to the magnetization due to the surface pinning field.

(ii) Site pinning field.

Here we consider a pinning field coupled to a single site in  a $D$-dimensional classical lattice model, or a pinning field in a $d$-dimensional quantum model, whose contribution to the action is, in the continuum limit,
\begin{equation}
h_0\int d^d\vec{x}d\tau\delta^d(\vec{x})\delta(\tau)\phi(\vec{x},\tau),
\end{equation}
where $\tau$ denotes the imaginary-time coordinate.

In this case a simple generalization of the argument of Sec.~\ref{sec:fss:h0} gives the scaling dimension of the pinning field $y'=y_h-D=-\beta/\nu$. Such value is always negative, therefore, at variance with the previous cases, the pinning field is an irrelevant perturbation, and the  RG flow  always leads to the $h_0=0$ fixed point. Since at this fixed point the system is translationally invariant, there are no additional irrelevant operators, beside the usual bulk ones. At the critical point, the magnetization exhibit the following scaling behavior [compare with Eqs.~(\ref{mH}) and (\ref{mBC})]
\begin{equation}
\begin{split}
m=&(L+c)^{-\beta/\nu}f_m(h_0(L+c)^{y'}, \{u_i(L+c)^{-\omega_i}\}, \rho)\\
&+ \frac{1}{L^D}g_{\rm site}(h_0), \qquad y'=-\beta/\nu.
\end{split}
\label{m_sitepf}
\end{equation}
Expanding Eq.~(\ref{m_sitepf}) for $L\rightarrow\infty$, and using the fact that for every lattice size $m=0$ when $h_0=0$, we obtain
\begin{equation}
\begin{split}
m=&A(\rho)(L+c)^{-2\beta/\nu}\cdot\\
&(1+A_1(\rho)(L+c)^{-\beta/\nu} + A_2(\rho)(L+c)^{-\omega}+\ldots)\\
&+\frac{1}{L^D}g_{\rm site}(h_0), \quad \omega=\text{min}\{\omega_i\}.
\end{split}
\label{m_sitepf_exp}
\end{equation}
Eq.~(\ref{m_sitepf_exp}) shows that the leading scaling exponent for the magnetization is, unlike the cases discussed above, $2\beta/\nu$.
In Appendix \ref{sec:Sn} we provide a proof of this statement based on a rigorous identity holding for classical $O(N)$-symmetric systems.
In Eq.~(\ref{m_sitepf_exp}), the exponent of the leading scaling correction is $\text{min}\{1,\{\omega_i\},\beta/\nu,D-2\beta/\nu\}$. For the three-dimensional $O(N)$ UC, the dominant correction-to-scaling exponent is given by $\omega=\beta/\nu\approx 0.5$ and corresponds to the next-to-leading Taylor expansion of the scaling function $f_m$ of Eq.~(\ref{m_sitepf}) in the first variable. In this case the background scaling correction does not play a relevant role, since it decays with an effective exponent $D-2\beta/\nu\approx 2$, for the 3D $O(N)$ UC.

Away from the bulk critical point, and in the para\-magnetic phase, the magnetization approaches quickly the thermodynamic limit of $0$, because the contribution to $m$ due to the local nonzero magnetization around the pinning-field site is $\propto L^{-D}$. In the ordered phase, the Gibbs weight for a configuration with $m$ antiparallel to $h_0$ is depressed by a factor ${\rm exp}(-Ch_0)$ with respect to a configuration with $m$ parallel to $h_0$, with $C$ a constant of $O(1)$ for $L\rightarrow\infty$. Therefore, in the broken phase the measured magnetization approaches a nonzero, but $h_0$-dependent value $m\propto(1-{\rm exp}(-Ch_0))$. Although this implies that the site pinning field can in principle be used to identify the ordered phase, the computed $m$ might be numerically small, thereby hindering a precise location of the critical point.

Beside these generalizations in the pinning-field geometry, the present study lends itself to further extensions. A numerical investigation of a classical lattice model in the three-dimensional $O(3)$ UC would be desirable, in order to compute the scaling dimension of the leading operator in the short-distance expansion of the order parameter and to verify if, as for the Ising UC, this operator is responsible for the slowly decaying scaling corrections discussed in Sec.~\ref{sec:results:heisenberg}. Moreover, a generalization to the 3D $O(N)$ UC, $N\ne 1, 3$ represents a natural extension of the present study. For these investigations improved lattice models like the Blume-Capel model considered here would be preferable, since the absence of relevant bulk scaling corrections then allows a precise determination of the critical exponents. Finally, another interesting issue is the crossover behavior of the order-parameter profile for small values of the pinning field.

%%%%%%%%%%%%%%%%%%%%%%%%%%%%%%%%%%%%%%%%%%%%%%%%%%%
\begin{acknowledgments}
F. P. T. is grateful to S. Dietrich for suggesting Ref.~[\onlinecite{Hanke-00}].
The allocation of CPU time within JARA-HPC at RWTH Aachen und  JSC J\"ulich and at LRZ  M\"unich is gratefully acknowledged.
F. P. T. is supported by the German Research Foundation (DFG)  DFG-FOR 1162  (AS 120/6-2).
F. F. A.  is supported by the German Research Foundation (DFG), under the DFG-SFB 1170 ToCoTronics (Project C01).
S.W. is supported by the German Research Foundation (DFG), under Grants WE 3649/3-1, DFG-RTG 1995,  and DFG-FOR 1807.
\end{acknowledgments}

\appendix

\section{FSS behavior of the local susceptibility at zero pinning field}
\label{sec:chiloc}
In this appendix we analyze the FSS behavior of the local line susceptibility defined in Eq.~(\ref{chilocdef}), in the absence of a pinning field. The results provide an alternative argument leading to Eq.~(\ref{yprime}).

We consider for illustration the classical lattice model and compute $\chi_{\rm loc}(\vec{x}_\perp,K,h_0,L_z,L)$ for $h_0=0$. The right-hand side of Eq.~(\ref{chilocdef}) can be computed using the scaling behavior of the two-point function which, according to standard scaling arguments and neglecting scaling corrections, is given by
\begin{equation}
\begin{split}
\<S_{(x,y,z)}&S_{(x,y,z')}\>-\<S_{(x,y,z)}\>\<S_{(x,y,z')}\>\\
&=\frac{1}{L^{D-2+\eta}}f_2\left(uL^{1/\nu},\frac{z-z'}{L},\rho\right), \quad h_0=0.
\end{split}
\label{scal2}
\end{equation}
By inserting Eq.~(\ref{scal2}) into Eq.~(\ref{chilocdef}) one finds
\begin{equation}
\begin{split}
&\chi_{\rm loc}(\vec{x}_\perp,K,h_0=0,L_z,L)\\
&=\frac{1}{L^{D-2+\eta}} L_z  \sum_{z} \frac{1}{L_z} f_2\left(uL^{1/\nu},\frac{z}{L},\rho\right) + f_{\chi,back}(K)\\
&=L^{3-D-\eta}\int_0^1d\widetilde{z} \rho f_2\left(uL^{1/\nu},\widetilde{z},\rho\right) + O\left(\frac{1}{L^2}\right) \\
&\quad + f_{\chi,back}(K),
\end{split}
\label{chilocint}
\end{equation}
where $f_{\chi,back}(K)$ is due to the terms in the sum of Eq.~(\ref{chilocdef}) for which $|z-z'|<<L_z$, i.e., the nonuniversal short-distance behavior of the two-point function which does not obey to the scaling behavior of Eq.~(\ref{scal2}); due to the translational invariance at $h_0=0$, $\chi_{\rm loc}$ is actually independent of $\vec{x}_\perp$.
On the other hand, $\chi_{\rm loc}(\vec{x}_\perp,K,h_0,L_z,L)$ can also be computed by differentiating twice the free energy density $\cal F$ with respect to $h_0$:
\begin{equation}
\chi_{\rm loc}(\vec{x}_\perp,K,h_0,L_z,L) = L^{D-1}\frac{\partial^2}{\partial h_0^2} {\cal F}(K,h=0,h_0,L_z,L).
\label{chilocdefder}
\end{equation}
Using Eq.~(\ref{fnonsingBC}) and Eq.~(\ref{fsingBC}) in Eq.~(\ref{chilocdefder}), and setting $h_0=0$ we find
\begin{equation}
\begin{split}
&\chi_{\rm loc}(\vec{x}_\perp,K,h_0=0,L_z,L) = \\
& L^{2y'-1}f_{\chi,\rm line}(uL^{1/\nu},\rho)
+ \frac{\partial^2f_{\rm line}^{(ns)}(K,h=0,h_0)}{\partial h_0^2}\Bigg|_{h_0=0},
\end{split}
\label{chilocder}
\end{equation}
where for the sake of clarity we have introduced a scaling function $f_{\chi,\rm line}(uL^{1/\nu},\rho)$ and, as in Eq.~(\ref{chilocint}), corrections to scaling have been neglected. By comparing Eq.~(\ref{chilocder}) with Eq.~(\ref{chilocint}) we finally recover Eq.~(\ref{yprime}).

\section{Pinning field in the Hubbard model on the honeycomb lattice}
\label{sec:honeycomb}
In Ref.~[\onlinecite{AH-13}] the quantum critical behavior of the Hubbard model on the honeycomb lattice was investigated by using the pinning-field approach. As a function of the hopping parameter $t$ and the Hubbard repulsion $U$, the Hubbard model on the honeycomb lattice exhibits a second-order quantum phase transition at $T=0$ between a semimetallic and an antiferromagnetic state, in the so-called Gross-Neveu Heisenberg UC~\cite{HJV-09}.
By extrapo\-lating the magnetization to the thermodynamical limit, Ref.~[\onlinecite{AH-13}] located the critical point at $U/t=3.78$, a value in line with recent numerical simulations~\cite{PTHAH-14,OYS-16}. According to the present study, the pinning field is a relevant perturbation for the line critical behavior, so that for $h_0\ne 0$, under the RG the model flows away from the ``ordinary'' fixed point $h_0=0$. In the absence of other line fixed points, a natural hypothesis is to assume that the RG flow leads to the $h_0=h_0^*=\infty$ fixed point, i.e., to a critical adsorption fixed point. Such a statement should be carefully checked, for instance by means of field theory calculations.

There are also some important quantitative differences with respect to the models studied here. The RG dimension of the pinning field at the $h_0=0$ fixed point is considerably smaller than for the 3D $O(N)$ UC. Using the results of Ref.~[\onlinecite{PTHAH-14}] in Eq.~(\ref{yprime}), we find $y'=0.15(8)$, with the results of Ref.~[\onlinecite{OYS-16}] we obtain $y'=0.25(3)$. Here, we have assumed that the dynamical exponent $z=1$, as implied by the Gross-Neveu-Yukawa field theory~\cite{HJV-09}. Such a small value of $y'$ indicates a rather slow crossover, so that one needs very large lattice sizes in order to realize the line fixed point, which is presumably the critical adsorption fixed point. Moreover, at variance with the spin models considered here, it is not a priori clear how to tune the model as to realize the line critical adsorption fixed point. Indeed, the introduction of a local magnetic field is described by the interaction term given in Eq.~(\ref{pf_interaction}), with $S^z_{i_0}=n_{i_0,\uparrow}-n_{i_0,\downarrow}$. The limit $h_0\rightarrow +\infty$ leads to a complete localization of the charge on the pinning-field site in the ground state, such that the occupation numbers are fixed as $n_{i_0,\uparrow}=1$, $n_{i_0,\downarrow}=0$. In this subspace of the full Hilbert space, the matrix elements of the hopping terms between the site $i_0$ and any other nearest neighbor site are suppressed to zero. Therefore, there is no interaction between the pinning-field site and the rest of the lattice, and the system is equivalent to a Hubbard Hamiltonian with a missing site (the pinning-field site), together with a spin degree of freedom, polarized parallel to the pinning field. Only the isolated pinning-field site contributes to the magnetization, while in the rest of the lattice the symmetry remains unbroken and the magnetization profile is exactly vanishing outside the pinning-field site. Notice that the difficulty in tuning the model parameters to the critical adsorption fixed point does not imply that such a fixed point is unreachable to the RG flow. Indeed, the description of the critical behavior in terms of a Gross-Neveu-Yukawa field theory emerges only after a RG treatment of the model, where the relevant degrees of freedom are identified and the resulting renormalized coupling constants are in general nontrivial functions of the parameters of the original lattice model. As such, it is {\it a priori} not obvious for which parameters of the lattice model one realizes an effective field theory with an infinite line pinning field. Nevertheless, the spin models studied in this paper can be thought of as a proper lattice regularization of a scalar $\phi^4$ theory, so that the critical adsorption fixed point can  indeed be obtained by setting the pinning field $h_0=\infty$; the results presented in Sec.~\ref{sec:results} support this observation.

Irrespective of the pinning-field fixed point $h_0^*$, some of the conclusions discussed in Sec.~\ref{sec:fss} are independently valid. For small values of $h_0$, the magnetization $m$ exhibits the scaling behavior shown in Figs.~\ref{fig:scalingmBC}, \ref{fig:figbhscaling}. At the fixed point $h_0^*$, the same line of argument that leads to Eq.~(\ref{m_norm}) is still valid, so that at the bulk critical point $m\propto L^{-\beta/\nu}$. Finally, for finite values of $h_0$ the scaling function of $m$ at criticality approaches a constant for $L\rightarrow\infty$. This can be understood by the following heuristic argument, adapted from a similar argument concerning the tail of the order-parameter profiles~\cite{BE-85}. After an RG transformation of scale $b$, the magnetization $m$ at the bulk critical point is transformed as
\begin{equation}
m=b^{-\beta/\nu}f(h_0(b), L/b).
\end{equation}
If we choose the scale $b$ to be large enough, and the size $L$ sufficiently large such that $L/b \gg 1$, we can substitute $h_0(b)$ by its fixed-point value $h_0^*$ (possibly the critical adsorption fixed point $h_0^*=\infty$) and set the scale by fixing $L/b=c$, so that we obtain
\begin{equation}
m=(L/c)^{-\beta/\nu}f(h_0^*,c) \propto L^{-\beta/\nu}.
\end{equation}
In other words, unlike what is stated in the appendix of Ref.~[\onlinecite{AH-13}], the asymptotic FSS behavior of the magnetization is characterized by the usual $\beta/\nu$ exponent.

Due to the crossover behavior found for $h_0\ne 0$, a scaling analysis of the magnetization which ignores the RG flow of the pinning field would measure an effective magnetization exponent $\beta/\nu$ which, only for large values of $L$, approaches the correct value. As explained above, the small value of $y'$ may render effectively impossible to reach the asymptotic behavior. Instead, for small values of the scaling variable $\widetilde{h_0}=h_0L^{y'}$, a first-order Taylor expansion of the leading scaling function $f_{mc}$ in Eq.~(\ref{mcrit}), gives $m\propto L^{-\beta/\nu}h_0 L^{y'}=h_0L^{-\eta}$ (see the corresponding discussion in Sec.~\ref{sec:results:heisenberg}.). Thus, for small values of $h_0L^{y'}$ one observes an effective magnetization exponent which is in fact the $\eta$ exponent. For the Gross-Neveu Heisenberg UC, one has $\eta=0.70(15)$ and $\beta/\nu=0.85(8)$~\cite{PTHAH-14}. Since these two values are close to each other, the analysis of Ref.~[\onlinecite{AH-13}], which incorrectly ignored the RG flow of the pinning-field line, nevertheless, did not introduce a significant bias in the results.

\section{FSS behavior of the magnetization in the presence of a single-site pinning field}
\label{sec:Sn}
In this appendix we show that, in the presence of a single-site pinning field, the effective scaling exponent of the magnetization is $2\beta/\nu$, in agreement with Eq.~(\ref{m_sitepf_exp}). The following argument is based on an exact identity which holds for classical $O(N)$-symmetric systems in the symmetric phase and is a generalization of an identity proven for Ising-like systems in Sec. 2.3 of Ref.~[\onlinecite{Cardy-book}].

Let us consider a classical $O(N)$-symmetric spin model on a lattice, in the absence of any external fields. Upon fixing the origin at a site $0$, the two-point function $\<\vec{S}_0\cdot\vec{S}_i\>$ is given by
\begin{align}
\<\vec{S}_0\cdot\vec{S}_i\> &= \frac{1}{Z}\int d\mu(\vec{S}_0) C(\vec{S}_0,i)
\label{Snintegral}\\
C(\vec{S}_0,i)&\equiv\vec{S}_0\cdot\int \prod_{k\ne 0}d\mu(\vec{S}_k) \vec{S}_ie^{-\beta{\cal H}\left(\vec{S}_0,\{\vec{S}_k\}_{k\ne 0}\right)},\nonumber
\label{CSn}\\
Z &= \int d\mu(\vec{S}_0)\int \prod_{k\ne 0}d\mu(\vec{S}_k)e^{-\beta{\cal H}\left(\vec{S}_0,\{\vec{S}_k\}_{k\ne 0}\right)},
\end{align}
where $Z$ is the partition function, $d\mu(\vec{S}) = \delta(\vec{S}^2-1)d^N\vec{S}$ is the $O(N)$-symmetric measure for a $N$-component spin variable $\vec{S}=(S^{(1)},S^{(2)},\ldots,S^{(N)})$ and we have explicitly indicated the dependence of the Hamiltonian ${\cal H}\left(\vec{S}_0,\{\vec{S}_k\}_{k\ne 0}\right)$ on $\vec{S}_0$. In the $O(N)$-symmetric phase, the function $C(\vec{S}_0,i)$ defined in Eq.~(\ref{CSn}) is actually independent of the value of $\vec{S}_0$. To show this, we compute it for a different value $\vec{S}'_0$ of the spin at the origin. If ${\bf R}\in O(N)$ is the matrix such that $\vec{S}'_0={\bf R}\vec{S}_0$, we have
\begin{equation}
\begin{split}
C(\vec{S}'_0,i) = {\bf R}\vec{S}_0\cdot\int \prod_{k\ne 0}d\mu(\vec{S}_k) \vec{S}_ie^{-\beta{\cal H}\left({\bf R}\vec{S}_0,\{\vec{S}_k\}_{k\ne 0}\right)}.
\end{split}
\end{equation}
Performing a change of variables $\vec{S}_k\rightarrow {\bf R}\vec{S}_k$ in the previous equation and exploiting the $O(N)$-invariance of the Hamiltonian ${\cal H}\left({\bf R}\vec{S}_0,\{{\bf R}\vec{S}_k\}_{k\ne 0}\right)={\cal H}\left(\vec{S}_0,\{\vec{S}_k\}_{k\ne 0}\right)$, we obtain $C(\vec{S}'_0,i) = C(\vec{S}_0,i)$. In particular, we can choose to fix $\vec{S}_0=\vec{e}_1=(1,0,\ldots,0)$. By using the same change of variables in the integral for the partition function $Z$ entering in the denominator of Eq.~(\ref{Snintegral}), the integral over $d\mu(\vec{S}_0)$ drops out and we find
\begin{equation}
\<\vec{S}_0\cdot\vec{S}_i\> = \frac{\int \prod_{k\ne 0}d\mu(\vec{S}_k) S^{(1)}_ie^{-\beta{\cal H}\left(\vec{e}_1,\{\vec{S}_k\}_{k\ne 0}\right)}}{\int \prod_{k\ne 0}d\mu(\vec{S}_k)e^{-\beta{\cal H}\left(\vec{e}_1,\{\vec{S}_k\}_{k\ne 0}\right)}}.
\label{Snresponse}
\end{equation}
The right-hand side of Eq.~(\ref{Snresponse}) can be interpreted as the local magnetization at the site $i$ when the spin at the origin is fixed to $\vec{e}_1$, i.e., in presence of an infinitely strong site pinning-field coupled to $\vec{S}_0$ and parallel to $\vec{e}_1$. Therefore, Eq.~(\ref{Snresponse}) can be expressed as the following identity
\begin{equation}
\<\vec{S}_0\cdot\vec{S}_i\>_{h_0=0} = \<S^{(1)}_i\>_{h_0=\infty},\qquad \vec{h}_0\parallel \vec{e}_1,
\label{Snidentity}
\end{equation}
where we have emphasized that the two-point function on the left-hand side is calculated in the absence of symmetry-breaking fields. By summing over the lattice site $i$ in Eq.~(\ref{Snidentity}) we obtain a relation between the magnetization $m(h_0=\infty)$ in the presence of an infinitely strong site pinning-field and the susceptibility $\chi(h_0=0)$ in the absence of external fields
\begin{equation}
m(h_0=\infty) = \frac{1}{L^D}\sum_i\<\vec{S}_0\cdot\vec{S}_i\>_{h_0=0}=\frac{1}{L^D}\chi(h_0=0).
\label{Snmvschi}
\end{equation}
Eq.~(\ref{Snmvschi}) holds in particular in a finite volume at criticality. Neglecting for simplicity corrections-to-scaling, and employing the standard FSS behavior of $\chi$ which can be obtained by differentiating twice Eq.~(\ref{fsingBC}) with respect to $h$, we find
\begin{equation}
m \propto \frac{1}{L^D}L^{2-\eta} \propto L^{-2\beta/\nu}.
\label{Snmexp}
\end{equation}
Due to universality, the result of Eq.~(\ref{Snmexp}) holds also for any system whose critical behavior belongs to the $O(N)$ UC.

\bibliographystyle{apsrev4-1_custom}
\bibliography{francesco,stefan}
\end{document}